\documentclass[11pt,prd,preprintnumbers,amsmath,amssymb, superscriptaddress]{revtex4}
\pdfoutput=1
\usepackage{amssymb}
\usepackage{amsmath}
\usepackage{enumerate}
\usepackage{graphicx}
\usepackage{subfigure}
\usepackage{multirow}
\usepackage{hyperref}
\usepackage{color}
\usepackage{appendix}
\usepackage{url}

\def\be{\begin{equation}}
\def\ee{\end{equation}}
\def\ba{\begin{array}}
\def\ea{\end{array}}
\makeatletter

\def\be{\begin{equation}}
\def\ee{\end{equation}}
\def\ba{\begin{eqnarray}}
\def\ea{\end{eqnarray}}
\def\beas{\begin{eqnarray*}}
\def\eeas{\end{eqnarray*}}
\def\sla{\raise.15ex\hbox{$/$}\kern-.57em}

\newcommand{\GeV}{\text{ GeV}}
\newcommand{\TeV}{\text{ TeV}}
\newcommand{\pb}{\text{ pb}}
\newcommand{\fb}{\text{ fb}}

\begin{document}
\title{\Large\bf Supersymmetric Crevices: Missing Signatures of RPV at the LHC}

\author{Peter W. Graham}
\affiliation{Stanford Institute for Theoretical Physics, Department of Physics, Stanford University, Stanford, CA 94305}

\author{Surjeet Rajendran}
\affiliation{Stanford Institute for Theoretical Physics, Department of Physics, Stanford University, Stanford, CA 94305}

\author{Prashant Saraswat}
\affiliation{Stanford Institute for Theoretical Physics, Department of Physics, Stanford University, Stanford, CA 94305}
\affiliation{Maryland Center for Fundamental Physics, Department of Physics, University of Maryland, College Park, MD 20742}
\affiliation{Department of Physics \& Astronomy, The Johns Hopkins University, Baltimore, MD 21218}

\preprint{UMD-PP-014-006}


\begin{abstract}
Supersymmetry is under pressure from LHC searches requiring colored superpartners to be heavy. We demonstrate $R$-parity violating spectra for which the dominant signatures are not currently well searched for at the LHC.  In such cases, the bounds can be as low as 800 GeV on both squarks and gluinos. We demonstrate that there are nontrivial constraints on squark and gluino masses with baryonic RPV ($UDD$ operators) and show that in fact leptonic RPV can allow comparable or even lighter superpartners. The constraints from many searches are weakened if the LSP is significantly lighter than the colored superpartners, such that it is produced with high boost.  The LSP decay products will then be collimated, leading to the miscounting of leptons or jets and causing such models to be missed even with large production cross-sections. Other leptonic RPV scenarios that evade current searches include the highly motivated case of a higgsino LSP decaying to a tau and two quarks, and the case of a long-lived LSP with a displaced decay to electrons and jets. The least constrained models can have SUSY production cross-sections of $\sim \text{pb}$ or larger, implying tens of thousands of SUSY events in the 8 TeV data. We suggest novel searches for these signatures of RPV, which would also improve the search for general new physics at the LHC.
\end{abstract}

\maketitle

\pagebreak

\tableofcontents

\pagebreak

\section{Introduction and Synopsis} \label{sec:Intro}

With the discovery of the Higgs boson, the Large Hadron Collider (LHC) has completed the discovery of the Standard Model, and is now tasked with discovering what, if anything, lies beyond it. In the complicated hadronic environment at the LHC, the search for new physics requires assumptions about the way this physics manifests itself in the detector. Given the strong motivation for the existence of new physics at the scales probed by the LHC, it is important to ask if the LHC could have missed discoveries because of such assumptions. 

Two of the most potent signatures to discriminate new physics from the large QCD backgrounds at the LHC are missing transverse momentum (MET) and hard leptons. In frameworks such as supersymmetry (SUSY),  there are many models  (such as the Minimal Supersymmetric Standard Model or MSSM) where SUSY events are rich with these objects, enabling their discovery. However, these features are not a theoretical requirement of the SUSY algebra, but are rather the product of additional assumptions such as the imposition of $R$-parity and ``generic" expectations about the low energy spectrum. 

$R$-parity violation (RPV) can radically alter the collider phenomenology of SUSY by causing the lightest supersymmetric particle (LSP) to decay. It is well known that this can be achieved by breaking either baryon or lepton number (but not both) while preserving the longevity of the proton (the primary motivation for introducing $R$-parity)~\cite{Hall:1983id}. It has long been known that baryon number violating RPV can greatly reduce the collider bounds on supersymmetry (e.g. ~\cite{KaplanUDD, Carpenter:2008sy, Displaced, VanillaSUSY}). In fact until relatively recently there were no experimental searches that could robustly constrain colored superpartners as light as a few hundred GeV. The tight LHC constraints on $R$-parity conserving SUSY have therefore motivated significant theoretical work on baryonic $R$-parity violation, in particular attempting to address issues regarding its flavor structure and embedding in a UV model~\cite{Nikolidakis:2007fc,Csaki:2011ge, Ruderman:2012jd, Krnjaic:2012aj, Bhattacherjee:2013gr, Csaki:2013we, Franceschini:2013ne, Florez:2013mxa, Krnjaic:2013eta, DiLuzio:2013ysa, Csaki:2013jza} as well as the origin of the primordial baryon asymmetry in the face of rapid washout of baryon number~\cite{Cui:2012jh, An:2013axa, RompineveSorbello:2013xwa, Baryonecrosis}. Lepton number violating RPV has in comparison seen little interest; it is often assumed that it cannot lower the bounds on superpartners as it introduces additional charged leptons and/or neutrinos.


In this work we challenge some of these commonly held assumptions about RPV at colliders. In particular, we show that leptonic RPV can in fact greatly relax the collider constraints on SUSY, and identify certain classes of models that particularly excel at evading LHC bounds, are potentially even less constrained than identical spectra with baryonic RPV. Further, we illustrate that even in simplified models of baryonic RPV designed to allow superpartners to be as light as possible, there are still significant constraints on the masses of colored superparticles.


Although missing energy signatures can be avoided, $R$-parity violating SUSY can often still be constrained by searches that select events with many high-$p_T$ collider objects, due to the tendency of RPV decays to produce many jets and/or leptons. In this paper, we show that this expectation can fail if the decaying LSP is significantly boosted, which occurs naturally if it is relatively light. The decay products of a boosted object are collimated along the direction of the boosted particle. As a result of this collimation, the decay products fail various isolation cuts, potentially resulting in mis-categorization of the event. For example, in the boosted decay of the  LSP to a lepton and jets through a lepton number violating operator, the lepton may not be identified as an isolated lepton and
the event could disappear into the large background for jet events.
This would greatly weaken the bounds from searches 
in leptonic final states. Similarly,  searches for baryonic $R$-parity violation rely on a large multiplicity of final state jets. When the jets are produced from a boosted decay, the resulting collimation will result in a decrease in the number of identified jets, affecting the multiplicity assigned to the event and potentially weakening searches that rely on jet multiplicity (e.g.~\cite{ATLASUDDGluino, ATLASMultijet20}).  The decaying LSP will have a large boost if it is much lighter than the copiously produced colored superpartners. Thus, a change to even very weakly coupled elements of the SUSY spectrum can dramatically alter the signatures of supersymmetry.

The choice of $R$-parity violating couplings of course also determines the collider phenomenology. These can in principle have an arbitrary flavor structure, allowing for a variety of decay modes, including ones that do not respect the na\"{\i}ve $SU(2)$ symmetry between charged leptons and neutrinos. We will show that within this model space there exist scenarios which largely evade existing LHC searches. In particular, we will demonstrate simple spectra with decays rich in taus, which are currently not much better constrained than purely hadronic final states. We also discuss the possibility of displaced decays of the LSP, and find that this may eliminate most of the constraints on an LSP decaying to an electron and quarks.


The main goal of this paper is to highlight the qualitatively different collider signatures that can emerge from choices of superpartner spectrum and RPV couplings. In light of the absence of signals of new physics in traditional channels and the fact that the low energy spectrum of superpartners is governed by undetermined physics at the scale where supersymmetry is broken, it is important to develop experimental techniques with sensitivity to the new types of signatures proposed here. In the bulk of this work, section~\ref{sec:Models}, we discuss various simplified models of RPV SUSY and point out how the above themes limit the sensitivity of existing searches to certain types of spectra. We conclude (section~\ref{sec:Conclusions}) with a discussion of the implications of these results for the future discovery prospects of supersymmetry.

\section{Models}\label{sec:Models}

\begin{figure}
		  \centering
        \subfigure[]{
                \includegraphics[width=2.0in]{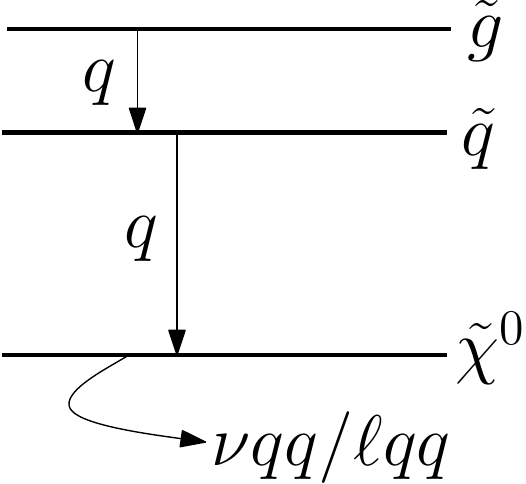}
                \label{fig:LQD_lnu_spectrum}
		  }
               \hspace{15mm} 
 	\subfigure[]{
                \includegraphics[width=2.4in]{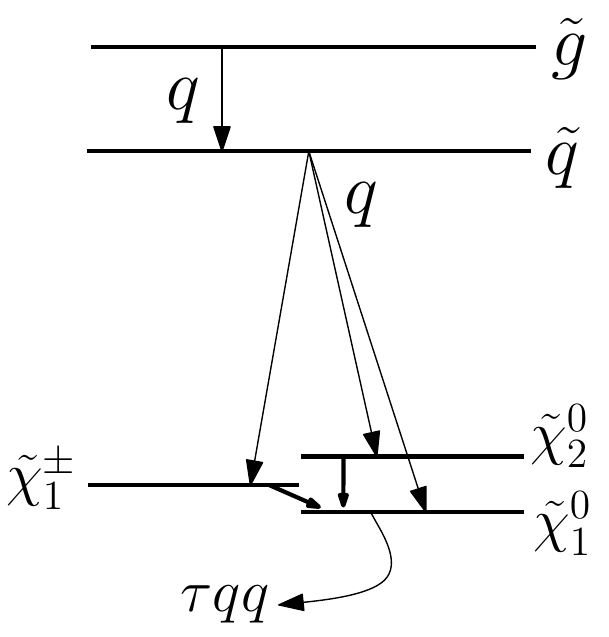}
                \label{fig:LQD_tauqq_nat_spectrum}
		  }
               \hspace{5mm} 
        \caption{Example spectra for the simplified models we consider. In (a), squarks and gluinos are produced and decay to a single neutralino LSP, which then can have various RPV decay modes that we explore. We take all of the squarks to be degenerate. In (b), there is a multiplet of nearly degenerate electroweakinos (e.g. higgsinos for small $\mu$ and large $M_1,M_2$); the colored superpartners can decay to any of these. Cascade decays within the multiplet will produce relatively soft particles which will be mostly irrelevant for collider phenomenology; in our event generation we take the splittings within a multiplet to be $< 1 \GeV$ to approach the limit where these soft objects can be ignored. Note that the presence of the other electroweakinos can still affect the collider phenomenology, e.g. by allowing the sbottom to decay to a top plus chargino instead of bottom plus neutralino.}
\label{fig:Spectra}
\end{figure}

To assess the degree to which $R$-parity violation can relax the LHC constraints on colored superpartners, we define several simplified models of squark/gluino production and cascade decay with various RPV decay modes of the LSP, of the general form illustrated in figure~\ref{fig:Spectra}. We take all of the squarks to be degenerate in these models (at least approximately), including those of the third generation. (This is to be contrasted with studies of ``natural SUSY''~\cite{Brust:2011tb, Papucci:2011wy} which assume that the first and second generation squarks are decoupled, greatly decreasing the total SUSY cross-section but causing the gluino to decay very distinctively into third generation quarks.) We focus primarily on models without cascade decays of electroweakinos amongst each other (e.g. $\chi^\pm \to W^\pm \chi^0$), as the additional leptons from such decays can greatly enhance the constraints. For the case of $R$-parity violation through the $UDD$ operator this effect was demonstrated in~\cite{VanillaSUSY, Vestiges}. To avoid this we consider models where the only electroweakinos lighter than the squarks and gluinos are either a single neutralino (as in figure~\ref{fig:LQD_lnu_spectrum}) or a nearly degenerate multiplet of charginos/neutralinos (figure~\ref{fig:LQD_tauqq_nat_spectrum}), such that their cascade decays do not produce hard objects relevant to collider physics. In these cases the main handles on the SUSY signal are the extra jets and/or leptons from the actual $R$-parity-violating LSP decay.  

Our exploration of these types of RPV spectra is in contrast to earlier studies of RPV collider phenomenology assuming specific models for the RPV couplings and/or particular minimal spectra, including ``natural'' spectra~\cite{Brust:2012uf, Evans:2012bf, Berger:2012mm, Franceschini:2012za, Berger:2013sir, Durieux:2013uqa, Bai:2013xla, Duggan:2013yna, Allanach:2012vj, CoverageNatural, Bhattacherjee:2013tha, Evans:2013uwa}. By including all of the colored superpartners in our simplified models we aim to identify models in which RPV hides a large total SUSY production cross-section from LHC searches.

Note that the spectra of figure~\ref{fig:Spectra} represent different simplified models depending on whether the gluino is heavier than or lighter than the squarks. For example, if the squarks are lighter than the gluino, then when squarks are produced at the LHC they will decay to one quark plus a neutralino; however if the squark is heavier than the gluino then it will undergo a two-step cascade producing three quarks and a neutralino. Searches selecting for high jet multiplicity will be especially affected by this.

The visibility of the LSP decay products varies with the LSP mass and decay length. Light LSPs can be produced with large boost from squark/gluino decays, causing their decay products to be collimated. This can prevent leptons from being sufficiently isolated from other particles to be accepted in collider searches, and can cause jets from the LSP decay to merge, reducing the observed jet multiplicity. 

If the LSP has a macroscopic decay length, $c\tau \gtrsim \text{mm}$, then its decay products will have displaced tracks which don't point back to the primary vertex. Such tracks are rejected by most analyses at ATLAS and CMS, preventing proper reconstruction of leptons and eliminating the sensitivity of standard searches in leptonic final states~\cite{Displaced}. However, dedicated searches for displaced decays can place strong bounds due to the low backgrounds to these types of events. A limited number of searches have been performed so far in the LHC data; for the models we consider the most relevant is an ATLAS search for displaced decays to a muon and jets~\cite{ATLASDispMuon}, which will also have some sensitivity if the muon originates from a tau lepton. Displaced decays to electrons and jets, however, are currently not directly constrained by displaced vertex searches. 


A recent CMS search for long-lived particles decaying to jets~\cite{CMSDisplacedDijet} may have sensitivity to the whole range decays we consider when the LSP is long-lived, as unlike previous LHC searches it does not require leptons or photons in the event to trigger on before reconstructing displaced vertices. The search however targets the case of a long-lived particle decaying to exactly two jets and places cuts which operate under the assumption that summing the momenta of the two displaced jets reconstructs momentum of the long-lived particle. It will therefore have significantly reduced sensitivity to a model with a three-body displaced decay. It is beyond our means to reproduce the event selection of this analysis, so the bounds, if any, from this search are currently unknown.   

Displaced decays may also introduce constraints from searches in jets plus MET, as they can introduce spurious missing energy in two ways: mismeasurement of the momentum vectors of particles as a result of assuming that all particles originated from the beamline, and the vetoing of displaced muons whose momentum is then unaccounted for. These effects can be mostly avoided if the LSP decay length is a small fraction of the calorimeter radius so that the error in momentum measurement is small, and if no muons are produced in the displaced decay. We will assume this in the models we consider.  

To determine the constraints on all models, we perform Monte Carlo simulations of collider events using {\tt MadGraph}~\cite{MadGraph} to generate the initial pair production, {\tt PYTHIA6}~\cite{PYTHIA} to decay and shower all particles, and a modified version of {\tt PGS}~\cite{PGS} to simulate detector reconstruction. Because {\tt PYTHIA} uses a flat matrix element for all three-body decays, we weight all events with the appropriate matrix elements for three-body neutralino and gluino decays. We obtain squark/gluino production cross-sections at NLL accuracy using {\tt NLL-fast}~\cite{Beenakker:2011fu}. We compute the expected event yield for several of the most relevant searches for new physics at ATLAS and CMS and determine the constraints on parameter space.  Further details on our MC simulation, including validation against official ATLAS and CMS results, are given in Appendix~\ref{sec:Appendix}.

\subsection{Baryonic RPV} \label{sec:Baryonic}

We first consider turning on the baryon-violating operator $UDD$, such that the neutralino LSP decays to quarks. If no top quarks are produced in this decay, then the spectra we consider (figure~\ref{fig:Spectra}) give essentially all-hadronic final states. It has long been known that this is an extremely challenging scenario for LHC searches. Recently however both ATLAS and CMS have performed searches for SUSY in all hadronic final states~\cite{ATLASUDDGluino,ATLASUDDGluino5,CMSRPVGluino,CMSRPVGluino5}, relying on the high jet multiplicity of the SUSY signal to reject QCD background. Official model interpretations for these searches are provided only for models with pure gluino production, with all squarks decoupled. 

\newlength{\figsize}
\setlength{\figsize}{2.9in}

\begin{figure}
		  \centering
        \subfigure[]{
                \includegraphics[width=\figsize]{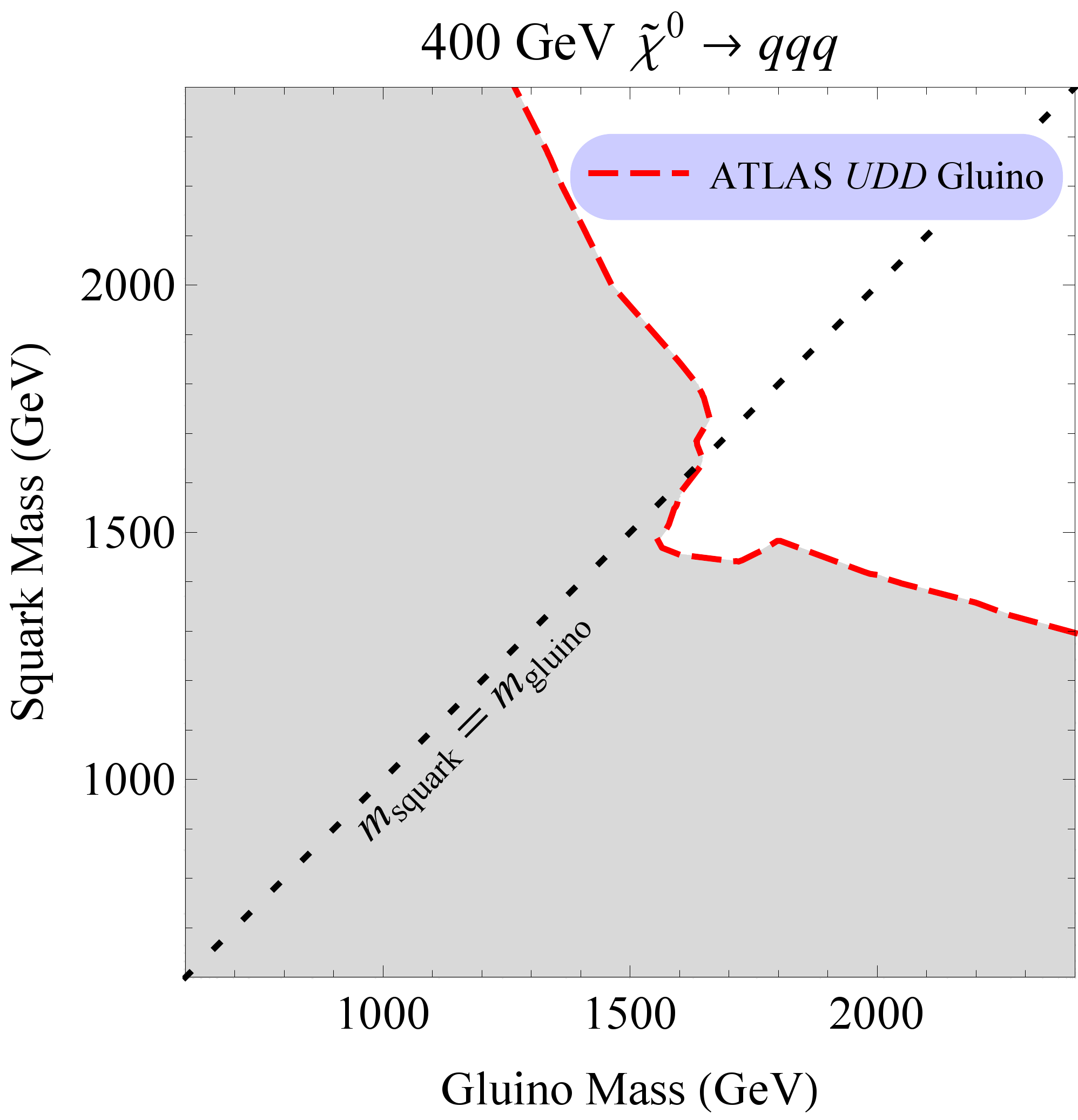}
                \label{fig:UDDchi_400}
		  }
               \hspace{3mm} 
 	\subfigure[]{
                \includegraphics[width=\figsize]{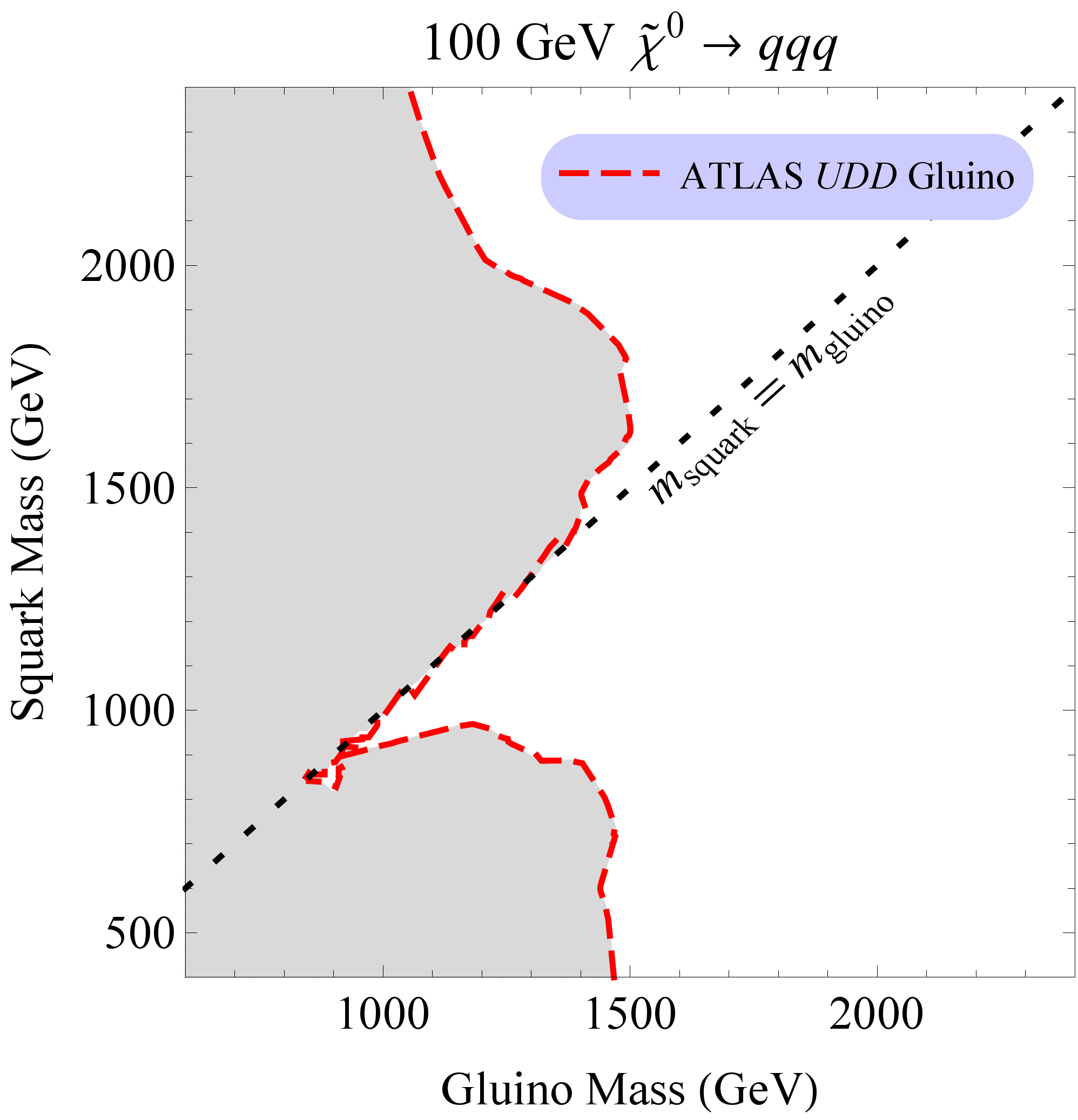}
                \label{fig:UDDchi_100}
		  }
       \caption{Constraints on a model of the type depicted in figure~\ref{fig:LQD_lnu_spectrum}, with a gluino, degenerate squarks, and a gaugino-like LSP decaying to three light quarks, for a 400 GeV LSP (a) and 100 GeV LSP (b). In both cases the constraint is derived from the ATLAS search for gluinos decaying through $UDD$~\cite{ATLASUDDGluino} (dashed red contour). Note the slightly different $y$-axis scales on the two plots.}
        \label{fig:UDD}
\end{figure}

We reinterpret the recent ATLAS search in $20 \fb^{-1}$~\cite{ATLASUDDGluino} for spectra including both squarks and gluinos, with all squarks degenerate. We take the neutralino to be gaugino-like; i.e. having equal coupling to the three generations of squarks. We intend to describe decays through $UDD$ operators, but this is complicated by the fact that our simulation pipeline cannot accommodate decays through operators with this color structure due to the limitations of {\tt PYTHIA6}. We approximate this model by replacing one of the quarks in the decay of the LSP with a ``fake quark": a very light, 1 GeV neutralino, which we decay to an electron and two quarks. Because the ``fake quark" is so light, it is produced with very high boost in collider events, such that its decay products are highly collimated in the lab frame. They will therefore appear as a single jet with momentum equal to the ``fake quark" momentum; the electron is non-isolated and thus simply adds to the calorimeter energy of the jet rather than being identified as a lepton object. The resulting collider phenomenology, we claim, is essentially identical to that of a true $UDD$ decay of the LSP-- the features not correctly modeled, such as the jet charge or the ratio of hadronic to electromagnetic calorimeter energies, are not immediately relevant to the searches we consider. We validate this modeling by reproducing the signal efficiencies reported by ATLAS in~\cite{ATLASUDDGluino}; see Appendix ~\ref{sec:ATLASUDDGluino_Validate} for details.

Our results are shown in figure~\ref{fig:UDD} as constraints in the plane of squark mass and gluino mass, for a spectrum of the type shown in figure~\ref{fig:LQD_lnu_spectrum} with two different choices of the LSP mass, 400 GeV (figure~\ref{fig:UDDchi_100}) and 100 GeV (figure~\ref{fig:UDDchi_400}). In most regions of parameter space the strongest bound is imposed by the most kinematically stringent signal region of~\cite{ATLASUDDGluino}, which requires 7 jets with $p_T$ greater than 180 GeV. Note that the constraints weaken significantly when the LSP is light, such that its decay products are boosted and tend to merge into a single jet rather than yielding high jet multiplicity. In fact, in this scenario initial and final state QCD radiation in signal events is essential in order to ever produce enough jets to pass the signal selection.

\begin{figure}
		  \centering

 	\subfigure[]{
                \includegraphics[width=\figsize]{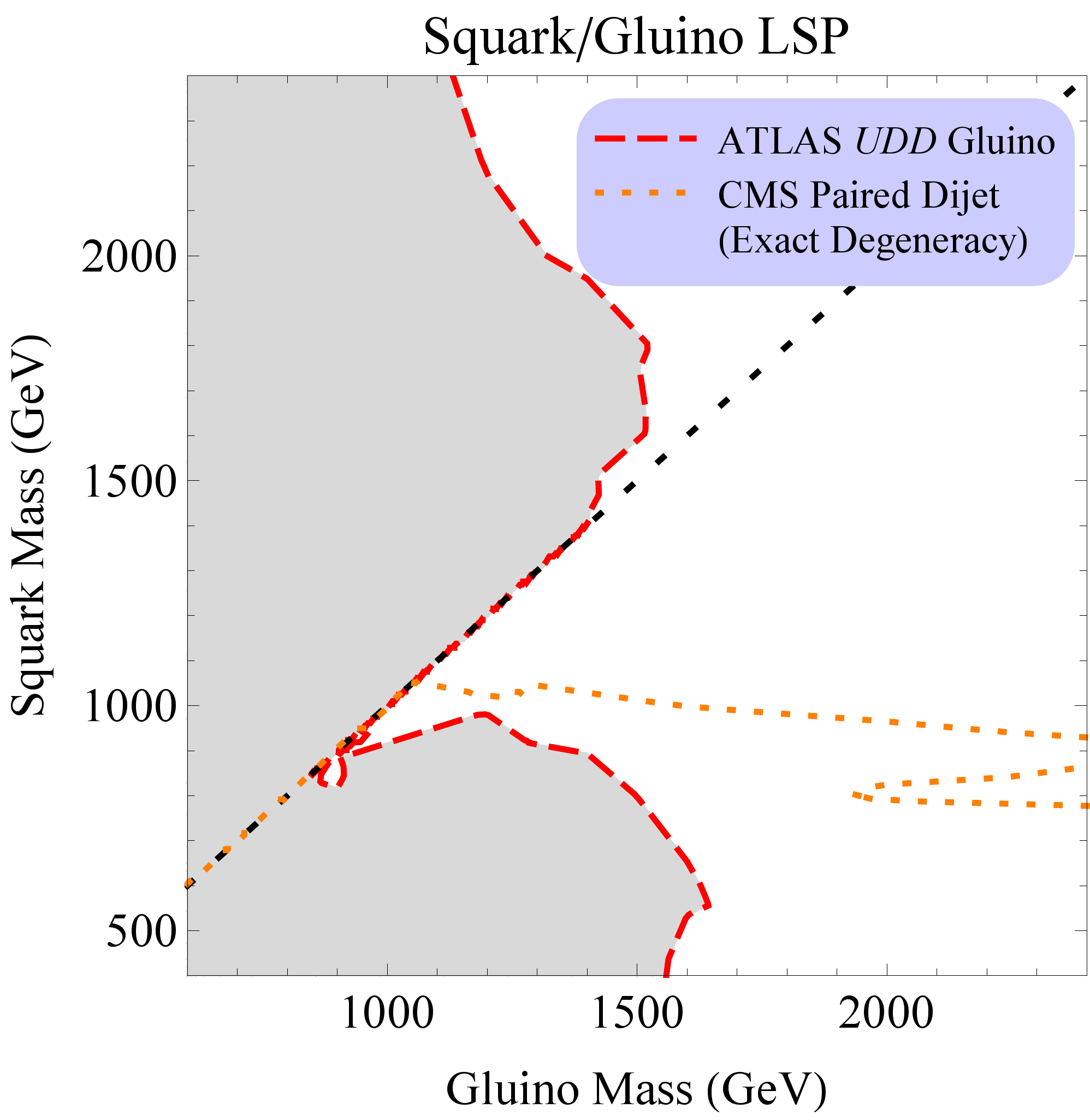}
                \label{fig:UDDnochi0}
		  }
          \caption{Constraints as in~\ref{fig:UDD}, but for a model without a neutralino, such that a squark or gluino is the LSP. The dashed red contour is again the constraint from the ATLAS search for gluinos decaying through $UDD$~\cite{ATLASUDDGluino}, while the dotted orange contour is a bound from a CMS search for paired dijet resonances~\cite{CMSDijetResonance}. The paired resonance bound applies when the squarks are exactly degenerate but will weaken greatly if the squark masses differ by $5-10\%$; since the exclusion is not robust we do not shade the region constrained by this search.}
        \label{fig:UDDnochi}
\end{figure}

One can also imagine a model without a neutralino LSP, with the lightest colored superpartner decaying directly through the $UDD$ operator. The constraints on such a model are plotted in figure~\ref{fig:UDDnochi}. The bound from the ATLAS RPV gluino search is essentially identical to those in the neutralino LSP model with a 100 GeV LSP (figure~\ref{fig:UDDchi_100}), confirming that in the latter case the neutralino LSP is typically boosted enough to look like a single jet. However, in the case of a squark LSP searches for paired dijet resonances~\cite{ATLASDijetResonance, CMSDijetResonance} can also be relevant. These searches select events with two pairs of dijets with similar dijet invariant mass, as would occur in squark pair-production with squarks decaying immediately to two light quarks. In our simplified model, the squarks are co-LSPs (if they are lighter than the gluino), so they essentially appear as a single ``effective" resonance with cross-section equal to the total squark pair production cross-section. The results of the CMS dijet resonance search~\cite{CMSDijetResonance} then imply a bound on the squark masses marked by the orange dashed line in figure~\ref{fig:UDDnochi}.

Of course, in a realistic SUSY spectrum the squarks will not all be exactly degenerate. (Although imposing flavor symmetries for the squarks for the first two generations is the simplest way to avoid bounds from flavor-changing neutral currents, this still allows different masses for the $\tilde{u}_R$, $\tilde{d}_R$, and $\tilde{q}_L$ fields.) Variation in masses of the squarks will appear to broaden the effective resonance; since the CMS dijet mass resolution is about $4.5\%$~\cite{CMSDijetResonance}, a variation of $5-10\%$ between the squark masses will cause them to no longer appear as a single true resonance but rather a much broader structure. Although such a broad peak may still be distinct from the QCD background, bounds on such a feature cannot be extracted from the limits on narrow resonances. Note in contrast that small variation between the squark masses will not dramatically affect the bounds from the ATLAS multijet search. We therefore do not shade the region constrained only by the dijet resonance bound in figure~\ref{fig:UDDnochi}, as this exclusion is not robust against small deviations from the simplified model. 

If the squark splittings are large enough and the RPV couplings weak enough, the heavier squarks may prefer to decay first to the squark LSP rather than directly to two quarks. The search~\cite{CMSDijetResonance} considers only dijet pairs formed out of the four highest-$p_T$ jets, so for large enough splittings the efficiency to reconstruct the LSP resonance is eroded, though the additional hard partons from the squark decays could contribute to bounds from the ATLAS multijet search.


For all models we find that the bound on the gluino becomes much stronger when the squarks are not decoupled. There is an apparent discontinuity across the black dashed lines in figures~\ref{fig:UDD} and~\ref{fig:UDDnochi} where the squark mass and gluino mass are equal; as discussed above the jet multiplicity increases greatly when the squarks are heavier than the gluinos, which strongly affects the signal efficiency in this search. Conversely, when the squark is just lighter than the gluino then the the quark from the gluino-to-squark decay tends to be soft, reducing the number of high-$p_T$ jets and relaxing the bound, resulting in the ``wedge'' of allowed parameter space just below the black dashed line. As the squarks become very heavy the limit on the gluino mass approaches the official ATLAS results for decoupled squarks, which we reproduce in Appendix~\ref{sec:Appendix} to validate our simulation. 

These results indicate that the LHC now places very significant constraints on the colored superpartners even in what was thought to be the best-case scenario for hiding supersymmetry. However, for a light LSP it is only gluino production that is actually constrained; light squarks can be allowed if the gluino is sufficiently decoupled. (We do not attempt to derive limits for squarks under 400 GeV, as at that point the expected number of SUSY events is so large that our finite Monte Carlo samples are not sufficient to estimate the extremely small signal efficiencies that could still result in constraints.) However, such a spectrum is hard to realize naturally since a large gluino mass tends to drag up the squark masses through the RGE's. Models with Dirac gluinos can avoid these large radiative corrections to the squark mass, but typically suffer from other model-building challenges~\cite{Vestiges}.

\subsection{LRPV without missing energy\texorpdfstring{ ($\chi \to \ell q q$)}{}}  \label{sec:NoMET}

In the rest of this work we will explore RPV with lepton number breaking. We will again consider simplified models of the form shown in figure~\ref{fig:LQD_lnu_spectrum}, with a gaugino-like LSP decaying to two light quarks and a charged or neutral lepton. We first focus on scenarios where the LSP decay always produces a charged lepton (rather than a neutrino). There are a few possibilities to ensure this decay mode:

\begin{itemize}
  \item Suppose that $L Q D$ operators are active and the sleptons are significantly lighter than the squarks, such that the LSP predominantly decays through off-shell sleptons. Then if the LSP is a neutral higgsino it will typically decay to a charged lepton, as it does not couple to the sneutrinos. Because of the hierarchy of Yukawa couplings, this will result in decays predominantly to tau leptons if the $L_i Q_j D_k$ couplings are comparable for $i = e, \mu,\tau$. The same conclusions hold if the LSP is a singlino that couples to the MSSM through mixing with the higgsinos.  

  \item One can introduce the non-renormalizable operator $\frac{1}{\Lambda} H_d Q U E$ into the superpotential. When the Higgs fields obtain their vevs this operator gives the desired three-body decay. It can be generated by integrating out a heavy lepton $L_4$ with a vectorlike mass and $R$-parity violating couplings, e.g.
\begin{equation}
W \supset M L_4 \bar{L}_4 + y^e_{4} H_d L_4 E + \bar{\lambda}' \bar{L}_4 Q_j U_k
\end{equation}
  \item Large left-right mixing in the slepton sector can cause the lightest charged slepton to be significantly lighter than the sneutrinos while still having appreciable left-handed component. In this case decays of the neutralino through $LQD$ will dominantly produce charged leptons even for bino or wino LSPs.  
\end{itemize}

\setlength{\figsize}{2.75in}

\begin{figure}
		  \centering
        \subfigure[]{
                \includegraphics[width=\figsize]{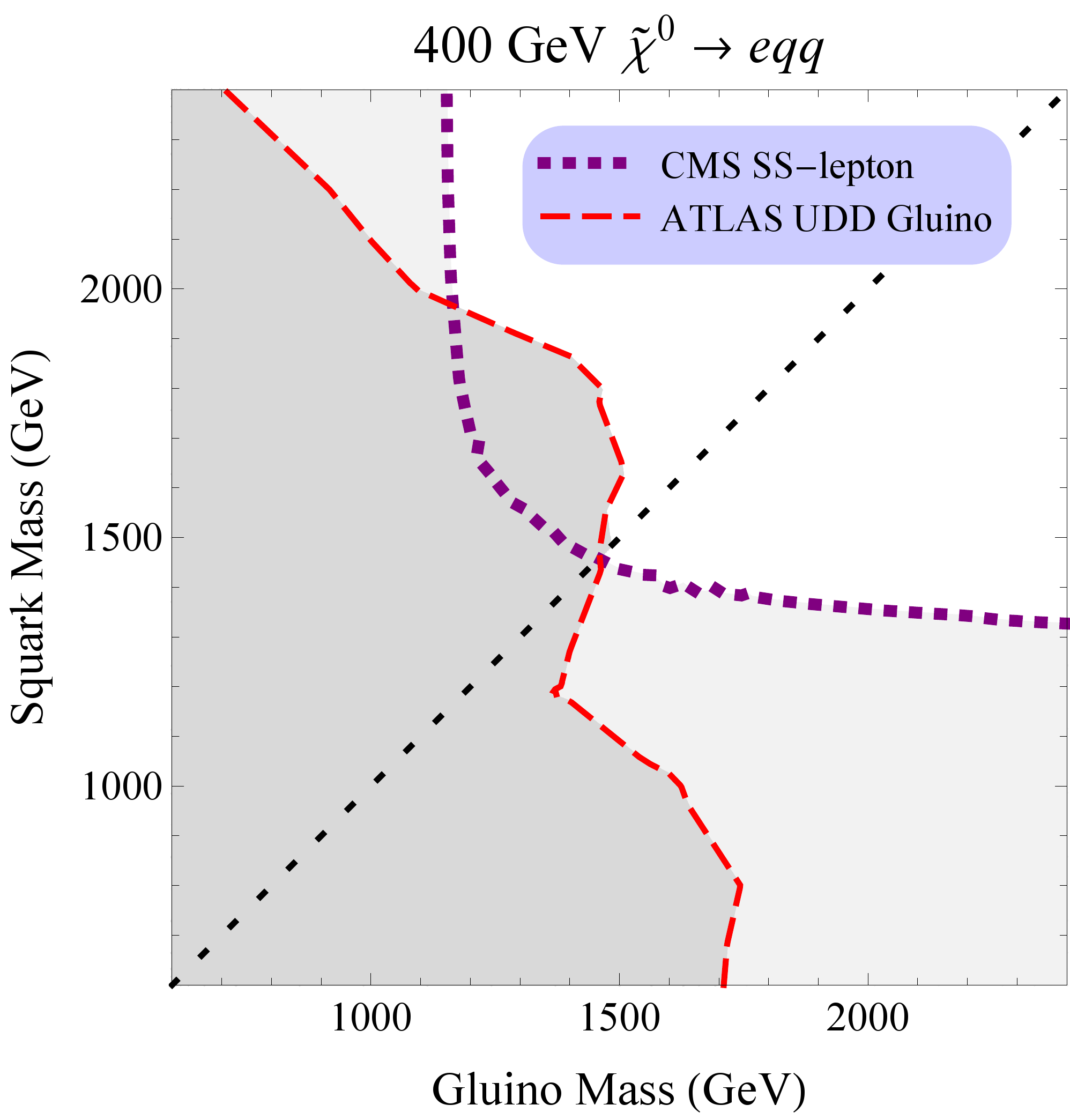}
                \label{fig:LQD_eqq_400}
		  }
               \hspace{3mm} 
 	\subfigure[]{
                \includegraphics[width=\figsize]{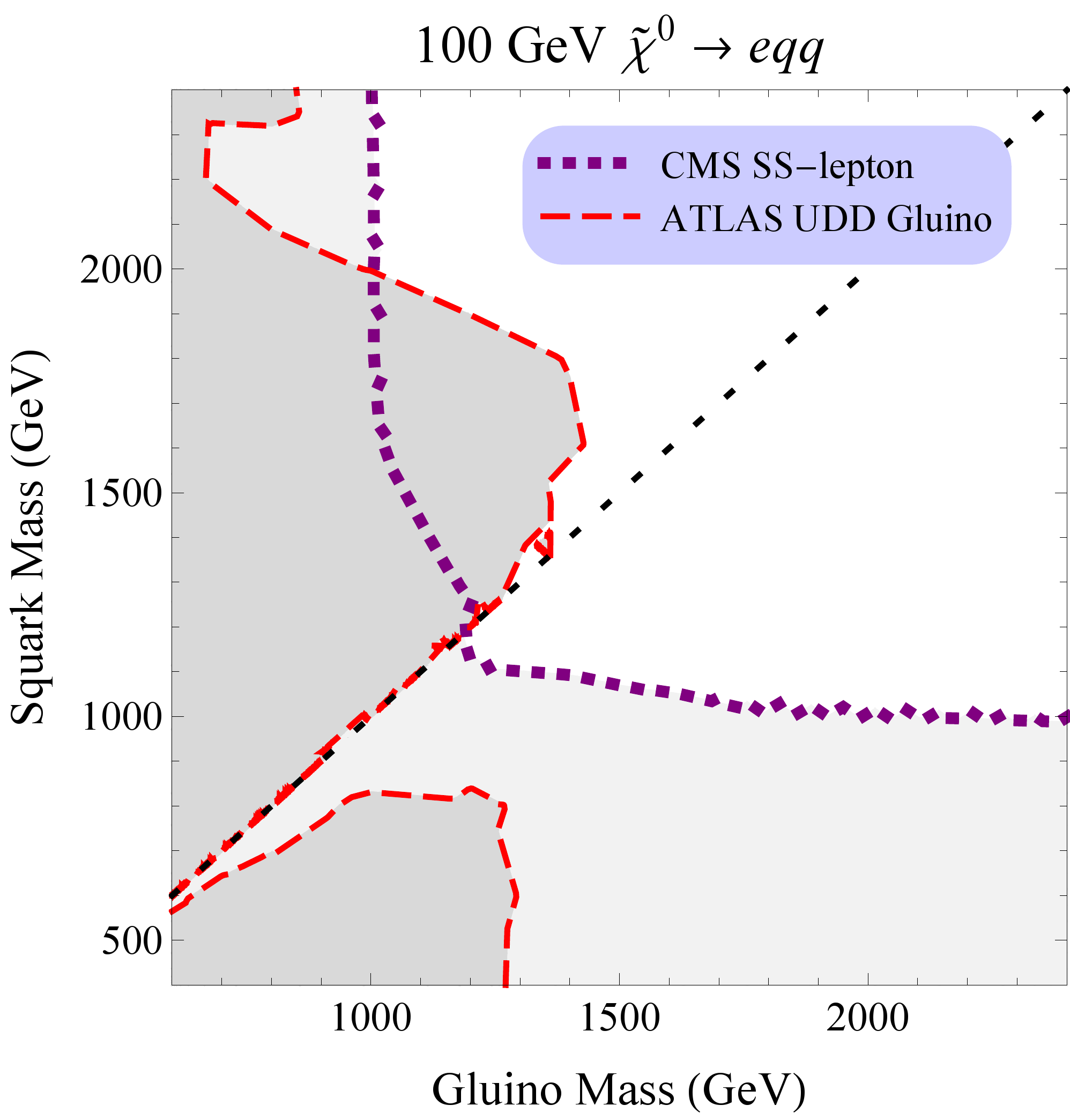}
                \label{fig:LQD_eqq_100}
		  }
        \subfigure[]{
                \includegraphics[width=\figsize]{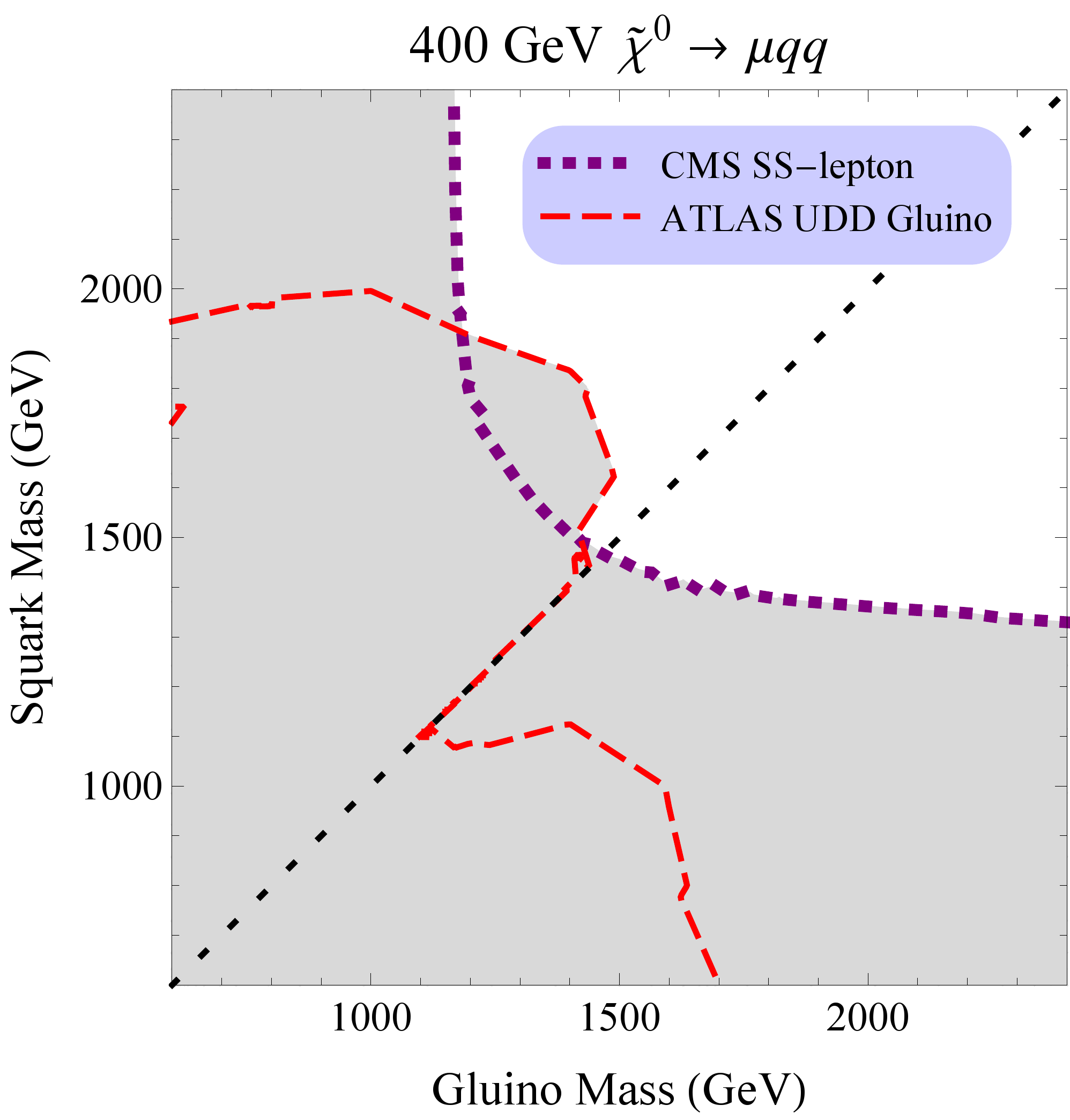}
                \label{fig:LQD_muqq_400}
		  }
               \hspace{3mm} 
 	\subfigure[]{
                \includegraphics[width=\figsize]{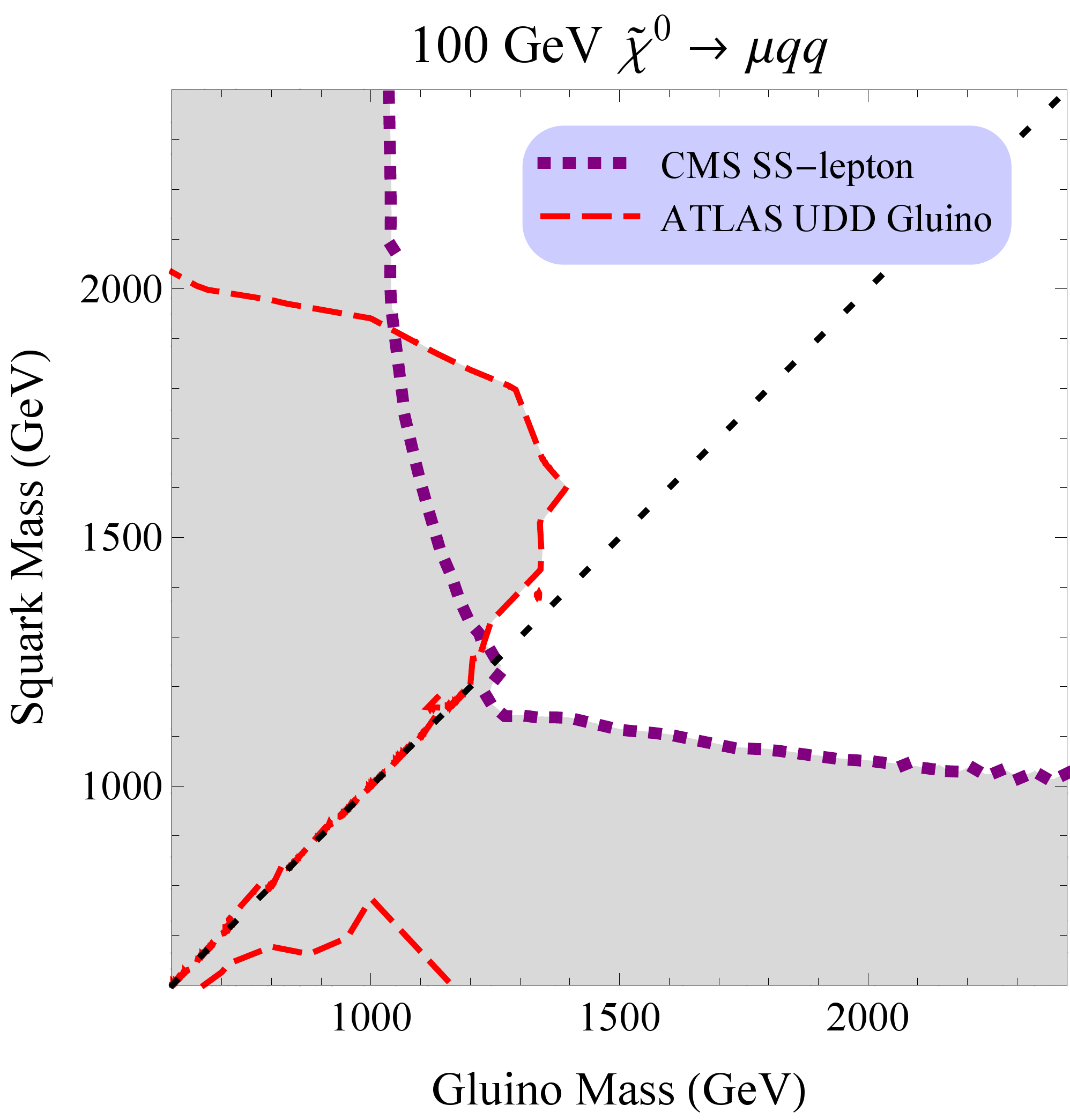}
                \label{fig:LQD_muqq_100}
		  } 
        \caption{Constraints on a mode of the type depicted in figure~\ref{fig:LQD_lnu_spectrum}, with a gluino, degenerate squarks, and a gaugino-like LSP decaying exclusively to $e q q$ (top plots) or $\mu q q$ (bottom plots). Two choices of LSP mass are shown here: 400 GeV (left plots) and 100 GeV (right plots); see figure~\ref{fig:LQD_emuqq_20} for the case of a 20 GeV LSP. Note the slightly different $y$-axis scales on the left-hand and right-hand plots. The relevant constraints are a CMS search for events with same-sign lepton pairs~\cite{CMSSSlep20} (dotted purple contour) and the ATLAS search for gluinos decaying through $UDD$~\cite{ATLASUDDGluino} (dashed red contour). The lighter shaded regions in~\ref{fig:LQD_eqq_400} and~\ref{fig:LQD_eqq_100} are only ruled out if the decay of the LSP is prompt; i.e.~displaced decays to electrons  allow for lighter superpartners.}
        \label{fig:LQD_emuqq}
\end{figure}

In this section we will focus on RPV decays to electrons and muons, in which case there is no true missing energy produced by the LSP decay. Decays to taus will be considered in section~\ref{sec:Tau-rich}. 

Figure~\ref{fig:LQD_emuqq} shows the bounds on such models in the cases where the lepton in the LSP decay is an electron (top plots) or muon (bottom plots), for LSP masses of 400 GeV (left plots) and 100 GeV (right plots). The purple dotted contours indicate the constraint from a CMS search in final states with same-sign leptons~\cite{CMSSSlep20}, which includes signal selections that do not require missing energy. The red dashed contours are once more the constraint from the ATLAS $UDD$ gluino search. We assume that direct production of the LSP is negligible and does not contribute to these constraints, which requires that the LSP be a bino, or, in extensions of the MSSM, another very weakly coupled state such as a singlino or photino~\cite{Photini}. In this case we must assume that the higgsinos and winos are heavier than the squarks and/or gluinos if we are to avoid $W$'s, $Z$'s and/or Higgs bosons from cascade decays. 

Note that once again constraints are much weaker when the LSP is light, such that its decay products tend to collimate. The efficiency of the same-sign lepton search is then eroded as the leptons tend not to be isolated from hadronic activity. The $UDD$ gluino search loses efficiency due to merging of the LSP decay products into a single jet. Since electrons deposit their energy in the calorimeters, they are included in jets if not identified as isolated objects. In contrast, muons contribute very little energy to calorimeter jets, resulting in weaker bounds from the $UDD$ gluino search in the case of LSP decay to muons.

In some CMS searches leptons are only checked for isolation from hadrons and photons, not other leptons; even if two leptons are extremely close to each other typically at least one is accepted~\cite{Matt}. Thus a highly boosted decay through an $LLE$ operator, for instance, would not evade leptonic searches at CMS.

If the decay of the neutralino is displaced, then the same-sign lepton search is not applicable, but as discussed above there are strong constraints from an ATLAS search for displaced vertices producing muons~\cite{ATLASDispMuon} and from jets + MET searches (as the muon energy is not counted). Displaced decays to electrons however are much less constrained by current searches than decays to muons, though the CMS search for displaced dijets~\cite{CMSDisplacedDijet} could be relevant. Therefore the light shaded region in figures~\ref{fig:LQD_eqq_400} and ~\ref{fig:LQD_eqq_100} should be taken as excluded only in the case of prompt LSP decays; for displaced decays we can only claim the bound form the ATLAS $UDD$ gluino search (dark shaded).  Thus displaced decays producing an electron could perhaps allow for much lighter colored superpartners. Also, in this case the LSP could be a higgsino or wino without running afoul of leptonic searches.

\begin{figure}
		  \centering
        \subfigure[]{
                \includegraphics[width=\figsize]{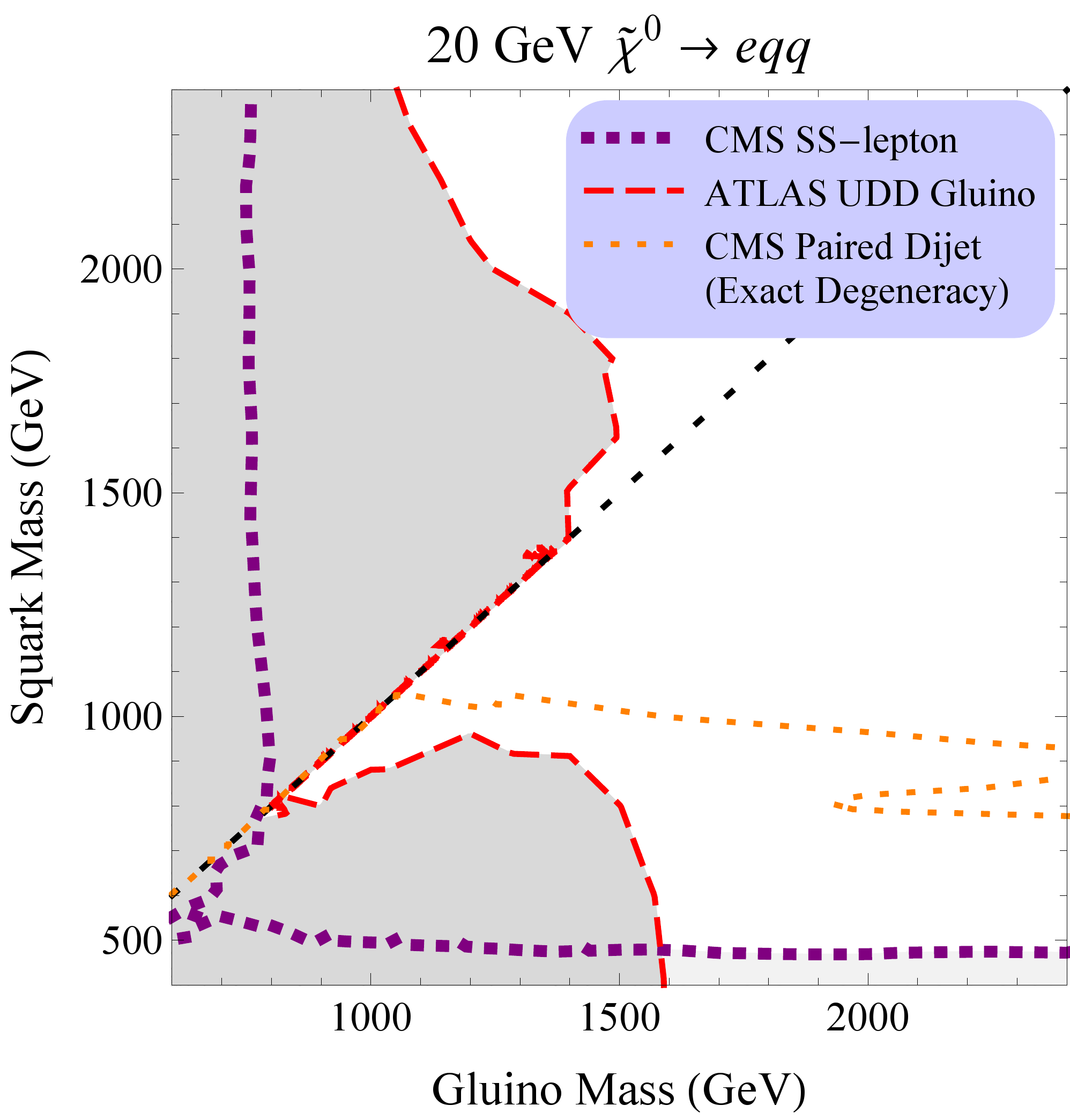}
                \label{fig:LQD_eqq_20}
		  }
               \hspace{3mm} 
 	\subfigure[]{
                \includegraphics[width=\figsize]{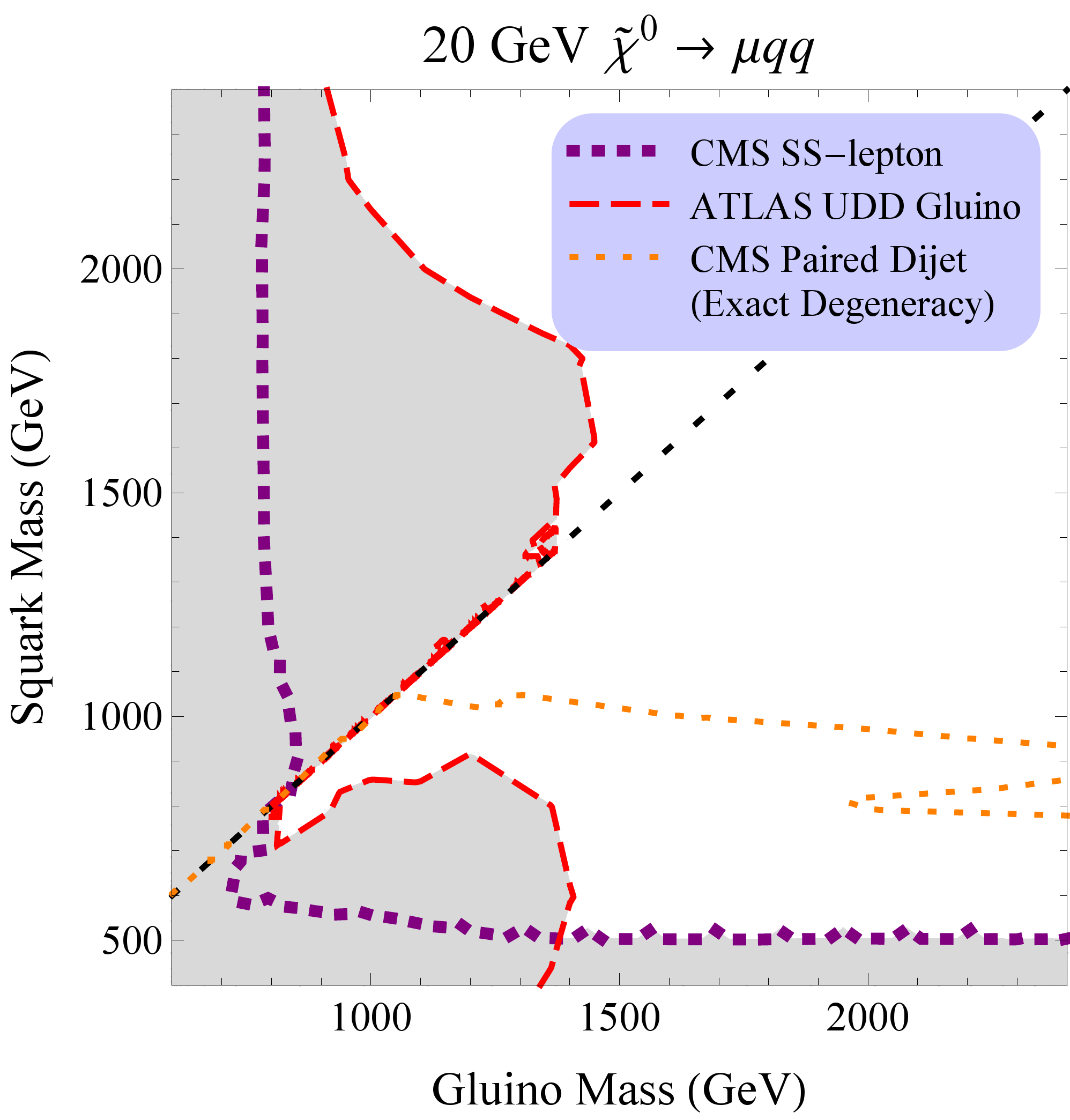}
                \label{fig:LQD_muqq_20}
		  }
\caption{Constraints on a model as in figure~\ref{fig:LQD_emuqq}, but with a 20 GeV LSP, again decaying to either $e q q$ (left) or $\mu q q$ (right). The constraints from the CMS same-sign lepton search~\cite{CMSSSlep20} are greatly weakened due to the high LSP boost. The dotted orange contour denotes a constraint from a CMS search for paired dijet resonances~\cite{CMSDijetResonance}, which applies when the squarks are exactly degenerate but  will weaken greatly if the squark masses differ by $5-10\%$; since the exclusion is not robust we do not shade the region constrained by this search.
}        
\label{fig:LQD_emuqq_20}
\end{figure}

These results suggest that all LHC constraints from searches for leptons could be avoided in models with very light LSPs that are produced with high boost, such that their decay products are collimated and the resulting leptons are no longer isolated from hadronic activity. A pure bino LSP cannot be produced directly at colliders and is allowed to be very light; likewise very weakly coupled states such as singlinos and photinos may naturally be light. Figure~\ref{fig:LQD_emuqq_20} shows the bounds on models with a 20 GeV LSP decaying to $e q q$ (left) or $\mu q q $ (right). The constraint from the same-sign lepton search is greatly weakened, to the point where it is in fact less sensitive than the ATLAS multijet search. However, since the LSP decay products almost always merge into a single jet in this scenario, the CMS search for paired dijet resonances~\cite{CMSDijetResonance} becomes relevant. The dotted orange contours of figure~\ref{fig:LQD_emuqq_20} are the constraints from this search on our simplified model with exactly degenerate squarks, very similar to those for squark LSPs decaying through $UDD$ (section~\ref{sec:Baryonic}). (Unlike ATLAS, CMS analyses include muons when clustering jets.) Once again though small perturbations away from exact squark degeneracy are enough to smear out the apparent resonance so that this search is not directly applicable. This is therefore not a robust constraint on the scale of the squark masses; light squarks ($500-1000 \GeV$) are allowed provided that they are split by $\sim 5-10\%$ of their mass. The merging of the LSP decay products into a single jet also means that the CMS displaced dijet search will have sensitivity if the decay is displaced (as it will appear to be two-body); in this case therefore decay to electrons will \emph{not} be less constrained when displaced. 

Remarkably, these models hide the colored superpartners as well or better than the best-case baryonic RPV scenarios, despite the presence of two leptons in every SUSY event. The colored superpartners could potentially all be lighter than $\sim \TeV$ in these models.


\begin{figure}
\includegraphics[width=3in]{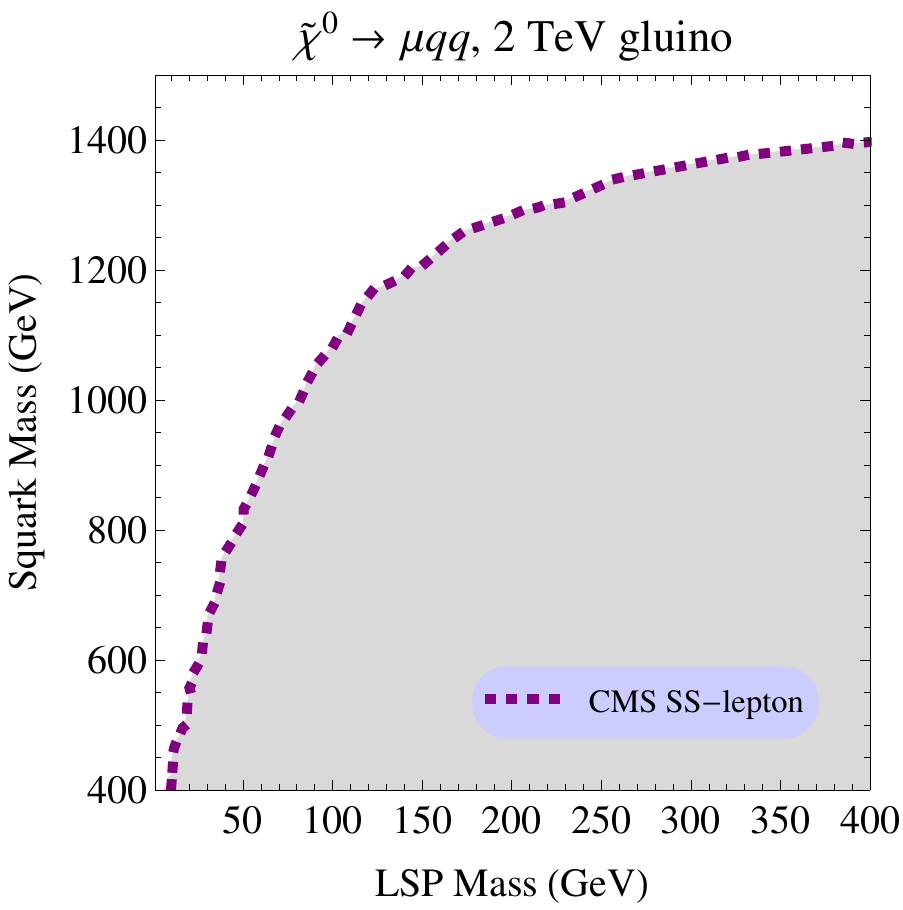}
\caption{Constraints on a model as in figures~\ref{fig:LQD_emuqq} and~\ref{fig:LQD_emuqq_20}, but shown in the plane of squark mass and LSP mass, with the gluino mass fixed to 2 TeV. The LSP is assumed to decay 100\% to a muon and two light quarks.  The effect of the collimation of the LSP decay products as the LSP becomes light is clear.
}         
\label{fig:Squark_vs_Chi}
\end{figure}

As in the case of baryonic RPV, in this boosted LSP scenario the squarks are allowed to be quite light if the gluino is sufficiently heavy. The dependence of the squark mass bound on the LSP mass is shown in figure~\ref{fig:Squark_vs_Chi}, which shows the constraints in the plane of squark mass and LSP mass for a fixed gluino mass of 2 TeV. This illustrates the continuous erosion of bounds from the same-sign lepton search with increasing LSP boost.

\begin{figure}
		  \centering
        \subfigure[]{
                \includegraphics[width=\figsize]{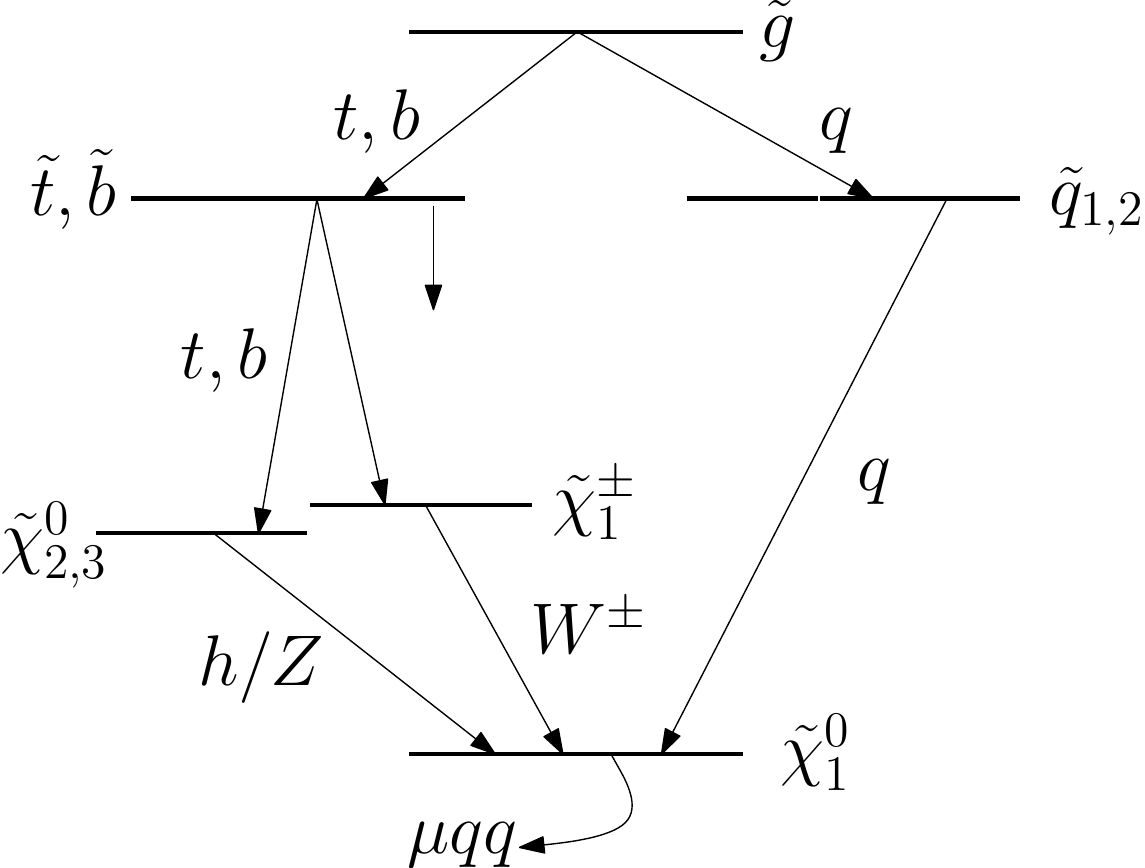}
                \label{fig:LQD_muqq_nat_spectrum}
		  }
               \hspace{3mm} 
 	\subfigure[]{
                \includegraphics[width=\figsize]{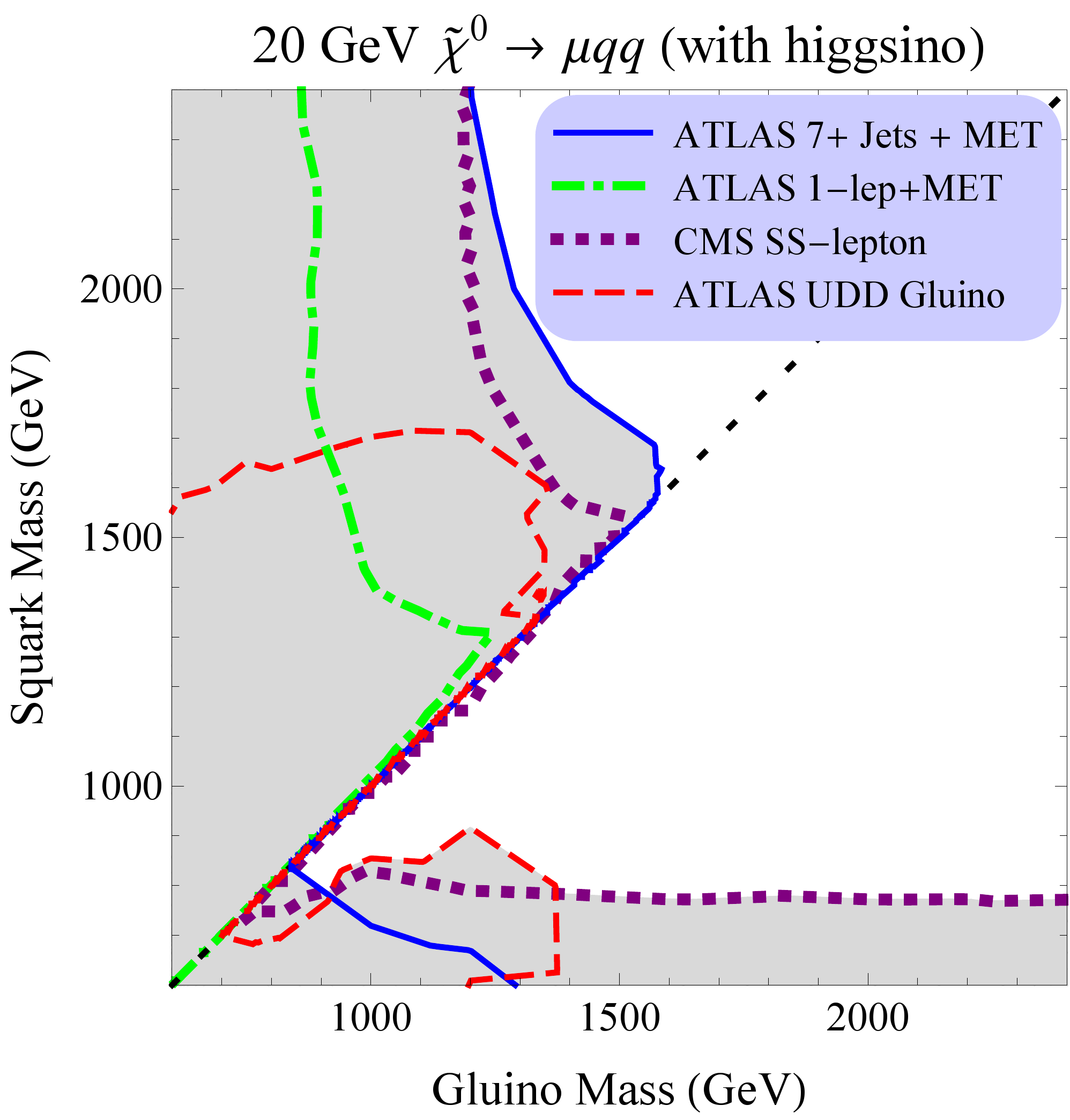}
                \label{fig:LQD_muqq_hino250_20}
		  }
\caption{ (a): A spectrum with a gluino, degenerate squarks, a degenerate Higgsino multiplet, and a bino-like LSP. If the gluino decays to light generation squarks, the decay cascade proceeds as in the spectra considered previously (figure~\ref{fig:LQD_lnu_spectrum}), with the light squarks decaying directly to the LSP. However, if the third-generation squarks are produced, then they tend to decay first to the higgsinos ($\tilde{\chi}^{2,3}_0$ and $\tilde{\chi}^\pm_1$), which then decay to the LSP through gauge or Higgs bosons. (b) Constraints on a model as in (a) with 250 GeV Higgsinos and a 20 GeV LSP. 
}         
\label{fig:LQD_muqq_hino250}
\end{figure}

In the above models the large higgsino mass (at least as heavy as the squarks) will dominate the electroweak fine-tuning, such that a fully natural theory cannot be achieved. It is interesting therefore to ask what occurs when we move away from the simplified model considered so far (figure~\ref{fig:LQD_lnu_spectrum}) and include higgsinos in addition to a very light bino-like LSP, as in figure~\ref{fig:LQD_muqq_nat_spectrum}. There are now many possibilities for the decays of the colored superpartners. We will assume that the relative couplings of the LSP and the higgsinos to the squarks are such that the heavy generations ($\tilde{t},\tilde{b}$) decay to the higgsinos, while the lighter generations decay directly to the LSP. The charged Higgsino will decay to a $W$ boson and the LSP; the neutral higgsinos can decay to either a Higgs boson ($h$) plus LSP and a $Z$ boson plus LSP. We will assume in this example that they decay to both final states with equal branching ratios. If the gluino is heavier than the squarks, it will decay to two third generation quarks and a higgsino. The constraints on this model in the squark-gluino plane are shown in figure~\ref{fig:LQD_muqq_hino250_20}, assuming 250 GeV Higgsinos and  20 GeV LSP decaying to $\mu q q$. The constraints on light squarks have tightened somewhat due to the presence of missing energy and non-isolated leptons in the decays of the third generation squarks. However, the bounds in the region of similar squark and gluino masses are comparable to the scenario of baryon number violation (figure~\ref{fig:UDD}); squarks and gluinos under a TeV are allowed, giving a large cross-section for the production of events with two non-isolated leptons.

We have seen that LRPV can allow colored superpartners (particularly squarks) to be much lighter than the current limits on $R$-parity conserving models. For some models, the bounds can in fact be comparable to or even weaker than the bounds on baryonic RPV. However, new search strategies could make significant progress. An obvious route is to simultaneously select on both of the distinguishing features of this signal, namely the presence of two leptons and the high jet multiplicity. A search in final states with same-sign leptons and many hard jets (without requiring MET) could further probe these models for LSPs above 100 GeV. It may be possible to attack the boosted case using an analysis with relaxed isolation requirements on leptons, relying on jet multiplicity cuts to control the background. The viability of such an approach depends on the Standard Model background for events with many hard jets and two high-$p_T$ leptons. The case of displaced decays could be probed by a search for displaced jets that does not assume specific kinematics as the CMS displaced dijet search does. Note that the SUSY production cross-section is quite large for some of the allowed model points, $\sim \pb$ (see figures~\ref{fig:Combined},\ref{fig:Combined20}), implying thousands of SUSY events in the 8 TeV data from which a signal could potentially be extracted.


\subsection{Tau-rich LRPV\texorpdfstring{ ($\chi \to \tau q q$)}{}}  \label{sec:Tau-rich}

We now turn to models in which the LSP decays to a tau lepton and two quarks. As discussed above, this is generic in the case of a neutral higgsino decaying through $L Q D$ operators unless those couplings involving taus are suppressed. A spectrum with squarks, gluinos and a higgsino LSP has the potential to be relatively natural. In order to accurately assess the prospects for this scenario, we will take the LSP to be higgsino-like in this section. 

\begin{figure}
		  \centering
        \subfigure[]{
                \includegraphics[width=3in]{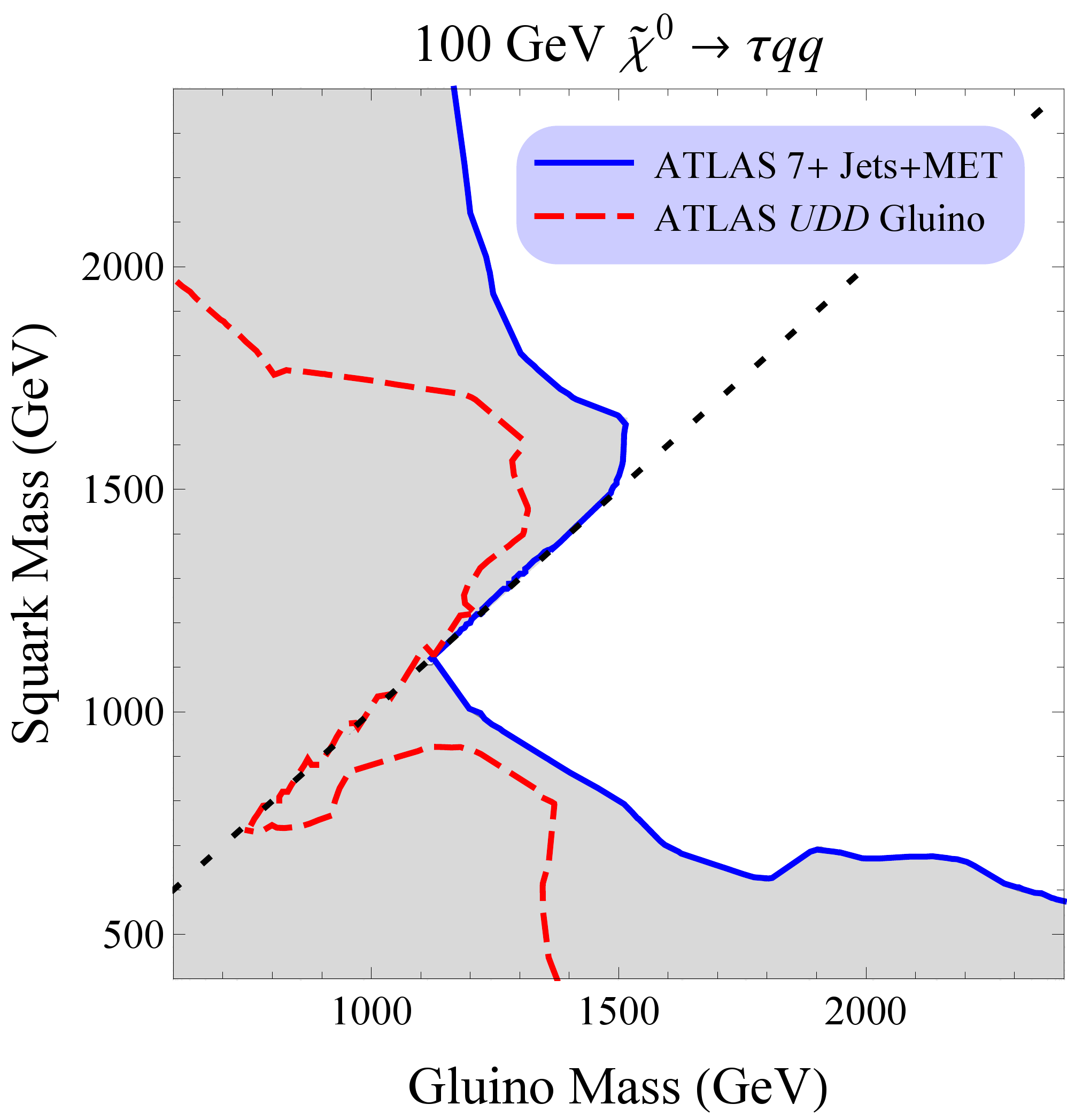}
                \label{fig:LQD_tauqq_nat_100}
		  }
               \hspace{1mm} 
          \caption{Constraints on a model of the type depicted in figure~\ref{fig:LQD_tauqq_nat_spectrum}, with a gluino, degenerate squarks, and a Higgsino LSP decaying to $\tau q q$. The relevant constraints in this case come from an ATLAS search for many (7+) jets and missing energy~\cite{ATLASMultijet20} (blue solid contour) and an ATLAS search for events with many high-$p_T$ jets~\cite{ATLASUDDGluino} designed to search for gluinos decaying through $UDD$ RPV operators (red dashed contour).}
        \label{fig:LQD_tauqq_nat}
\end{figure}

We first consider a spectrum of the form shown in figure~\ref{fig:LQD_tauqq_nat_spectrum}, with a nearly degenerate Higgsino multiplet. We take the splittings within the multiplet to be 1 GeV, such that intra-multiplet cascade decays do not produce additional collider objects. We decay the third-generation squarks and gluino appropriately depending on their dominant Higgsino coupling (e.g., we take left-handed sbottoms to decay 100\% to a top and a chargino). The most important difference from the gaugino-like scenario considered above occurs when the gluino is lighter than the squarks, in which case the gluino undergoes three-body decays preferentially to the heavy generations-- we take the gluino to decay 50\% to a neutralino and two top quarks and 50\% to a chargino (of either sign), a top quark, and a bottom quark.

In figure~\ref{fig:LQD_tauqq_nat}, we show the most relevant constraints on the squark and gluino masses assuming this spectrum with the LSP decaying 100\% to $\tau q q$, for a 100 GeV LSP. The blue contour indicates the constraint from an ATLAS search for many (7+) jets and missing energy~\cite{ATLASMultijet20}.  In this case the missing energy in these events arises mainly from neutrinos produced in tau decay, and is greatly suppressed compared to models with stable neutralinos. Unlike most LHC searches in jets and missing energy however, this search does not use use a MET trigger to collect the data sample but rather a multijet trigger, and merely requires that the MET in an event be statistically significant. The high jet multiplicity of these events in conjunction with moderate MET cut is sufficient to place constraints, particularly on the gluino mass; the squarks are much less constrained due to the lower jet multiplicity in squark production events. The red dashed lines are once more the constraint from the ATLAS $UDD$ gluino search~\cite{ATLASUDDGluino}.

We do not present bounds from searches selecting hadronic taus in the final state, as the algorithms used by ATLAS and CMS for tau reconstruction are not reproduced by our detector simulation. The relevant searches all require large missing energy however, so we do not expect them to place competitive bounds. We have checked that, using the default PGS tau reconstruction, the bounds from the ATLAS search in jets + taus + MET~\cite{ATLAStau} places much weaker bounds than the multijet + MET search.




It is interesting to note that in the region where the squark and gluino masses are similar, the constraints are comparable to those in the baryonic RPV case (figure~\ref{fig:UDD}). This suggests that there could be models with this form of lepton-number-violating RPV that are only as fine-tuned as baryonic RPV models with similarly light higgsinos. This would achieve a more natural electroweak scale without the complications of baryon number violation for baryogenesis, nucleon decay, etc. As in the case of LSP decays to light leptons, new searches could capitalize on the non-hadronic aspects of these SUSY events, though relying on hadronic taus is more challenging (and will likely be impossible for taus that overlap with jets). 


\begin{figure}
\includegraphics[width=3in]{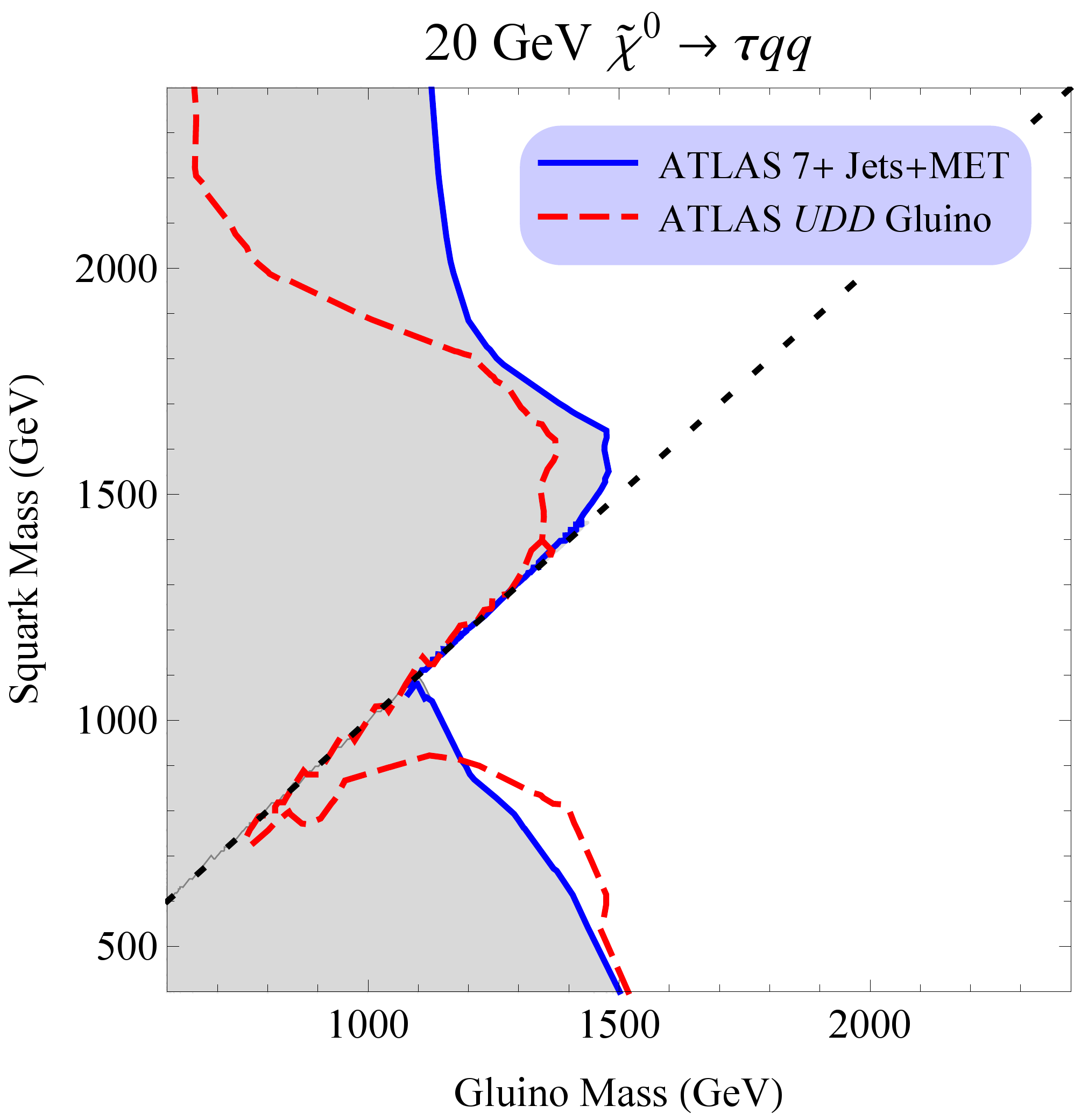}
\caption{Constraints on a model as in figure~\ref{fig:LQD_emuqq_20}, but with a 20 GeV LSP decaying to a tau and two light quarks.
}         
\label{fig:LQD_tauqq_nat_20}
\end{figure}




As before, we may also consider a very light LSP, though this cannot be a higgsino or wino due to LEP constraints. We therefore return to gaugino-like couplings when considering LSPs lighter than $\sim 100 \GeV$. In figure~\ref{fig:LQD_tauqq_nat_20} we show constraints on a model exactly as in figure~\ref{fig:LQD_emuqq_20} but with the LSP decaying to a tau and two quarks. A window for very light squarks exists in this case, as the boosted LSP prevents squark pair-production events from having high jet multiplicity. Unlike the case of LSP decay to the lighter leptons, here squarks will not appear as true dijet resonances due to the missing energy from the tau decay. Although we find that squark production in this model still gives a peak in the paired dijet mass spectrum (somewhat below the mass of the squarks), it is much broader than would be expected from a true resonance. The results from the CMS paired dijet search are therefore not directly applicable, even if the squarks are indeed exactly degenerate.

\subsection{Bilinear LRPV\texorpdfstring{ ($\chi \to \nu b b$)}{}}  \label{sec:Bilinear}

Next we consider the same squark-gluino-LSP spectrum, but with the LSP decaying 100\% to a neutrino and two bottom quarks. This decay is predicted by bilinear RPV, in which the dominant source of $R$-parity violation is the bilinear term $L_i H_u$ in the superpotential. After mass diagonalization the leptons and Higgs doublets mix by some angle $\epsilon_i$, giving rise to trilinear RPV terms of the form $\epsilon_i y^e_{jk} L_i L_j E_k$ and $\epsilon_i y^d_{jk} L_i Q_j D_k$ where $y^e$ and $y^d$ are the lepton and down-type quark Yukawa matrices respectively. The dominant operators are therefore $L_i Q_3 D_3$, which can allow the neutralino to decay to a charged lepton, top quark and bottom quark or a neutrino and two bottom quarks. If the LSP is lighter than the top quark, only the latter decay can occur. Decays through the operator $L_i L_3 E_3$ will generally be subdominant, but could give rise to lepton-rich signatures.

The phenomenology of this model in the light of 7 TeV LHC data was discussed extensively in~\cite{Displaced}. In particular, it was shown that neutrino mass constraints require the $R$-parity violating couplings in this model to be weak enough that the LSP tends to have a displaced decay, in which case the bounds on superpartner masses were drastically weakened. The macroscopic displacement of tracks prevents reconstruction of objects such as leptons and $b$-jets, or for CMS studies prevented the identification of jets completely. The only relevant constraints are therefore from ATLAS searches in final states with jets and missing energy. 

\begin{figure}
		  \centering
        \subfigure[]{
                \includegraphics[width=3in]{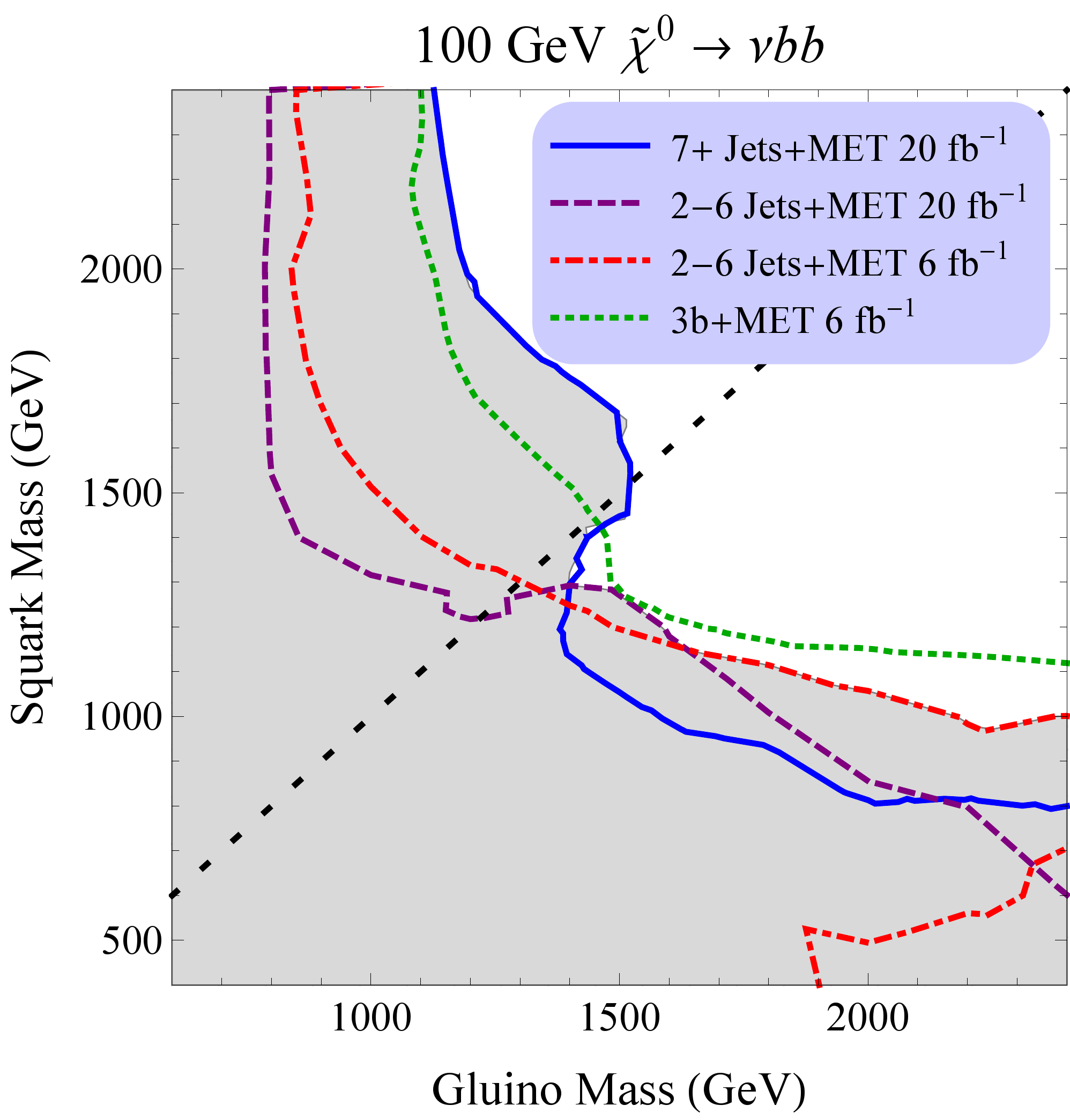}
                \label{fig:LQD_nubb_100}
		  }
               \hspace{1mm} 
	\subfigure[]{
                \includegraphics[width=3in]{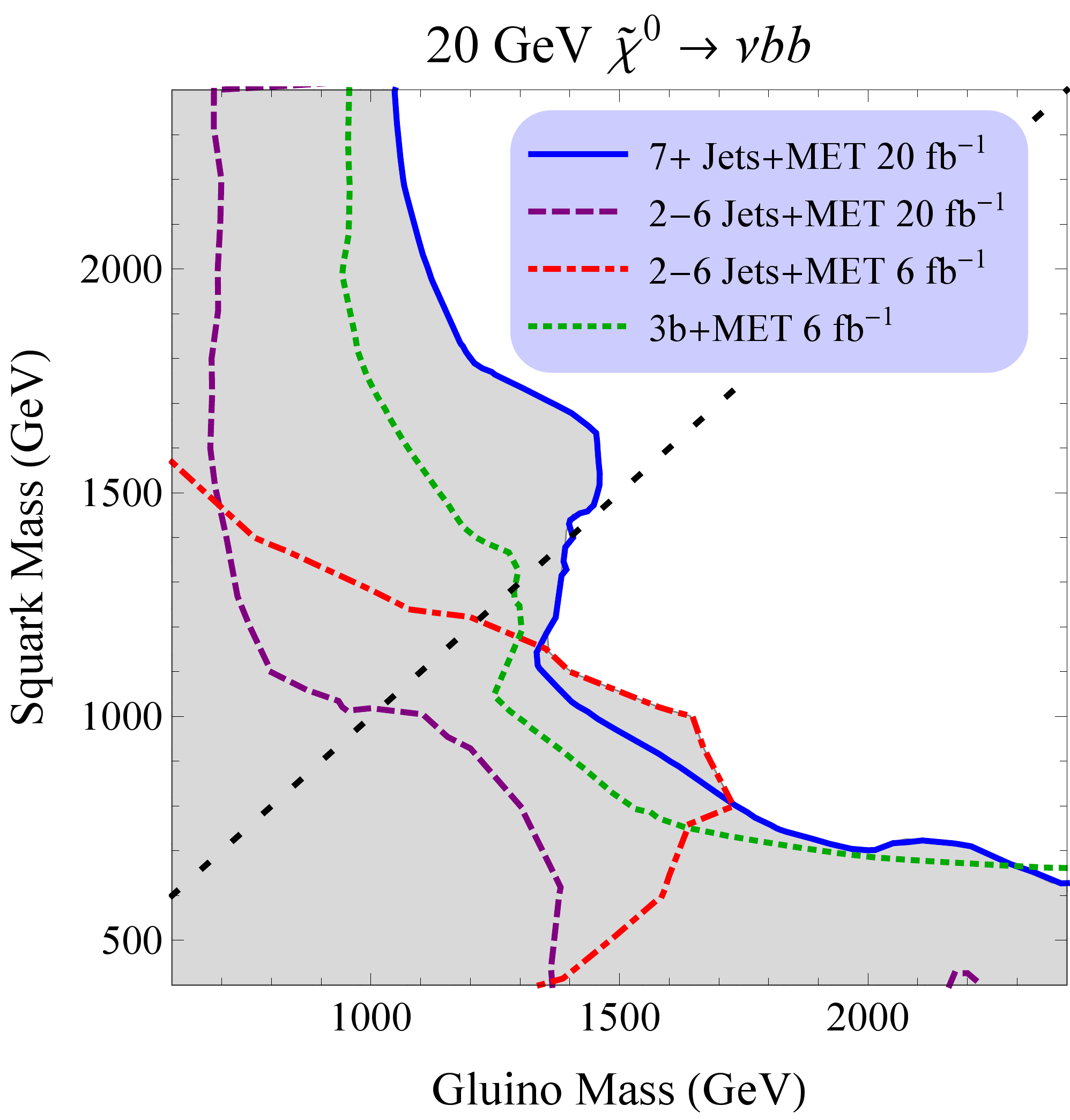}
                \label{fig:LQD_nubb_20}
		  }
        \caption{Constraints on a model as in figure~\ref{fig:LQD_emuqq}, but with the LSP decaying exclusively to $\nu b \bar{b}$. The strongest bounds come from the ATLAS search in many jets and missing energy. The constraints come from ATLAS searches in many (7+) jets and missing energy~\cite{ATLASMultijet20} and from two different analyses in final states with 2-6 jets and missing energy~\cite{ATLASJetsMET8,ATLASJetsMET20}. Of the latter, one analysis searches in the full 8 TeV data set of 20 $\fb^{-1}$~\cite{ATLASJetsMET20} while the other only analyzes 6 $\fb^{-1}$~\cite{ATLASJetsMET8}; however because of the different signal selections made, the latter places better bounds on this model in some regions of parameter space.}
        \label{fig:LQD_nubb}
\end{figure}

In~\cite{Displaced} it was found that searches in the 7 TeV data still allowed a ``window'' of very light squarks ($\sim 500 \GeV$). Here we update these results to account for the searches in the full 8 TeV data set. Figure~\ref{fig:LQD_nubb} shows the constraints on this simplified model for various values of the LSP mass. The shaded regions are the constrained parameter space assuming displaced decays, for which only ATLAS jets + MET searches place bounds. We find that the 8 TeV jets + MET searches have closed the window of light squark mass. We include the ATLAS searches in final states with 2-6 jets and missing energy for both $6 \fb^{-1}$~\cite{ATLASJetsMET8} and $20 \fb^{-1}$~\cite{ATLASJetsMET20} of 8 TeV data; the former gives stronger bounds in some cases due to the different signal region criteria used in the two analyses. In particular, for heavy gluinos and light squarks the 5-jet selection places the best constraints for both searches, with the $6 \fb^{-1}$ search placing a stronger bound due to the weaker constraint on the missing energy imposed in that analysis. For prompt decays, an ATLAS search in final states with missing energy and 3 or more $b$-tagged jets~\cite{ATLAS3b} places slightly stronger bounds on the squark mass for a sufficiently heavy LSP (green dotted line in figure~\ref{fig:LQD_nubb}).

If the LSP decay is displaced, then again the CMS search for displaced dijets could be relevant, though its sensitivity to three-body decays is unknown. Therefore we obtain conservative bounds on this model by assuming that the decay is displaced, such that the search in $b$-tagged jets + MET~\cite{ATLAS3b} doesn't apply, and that the CMS displaced dijet search places no bound regardless. This corresponds to the shaded regions in figure~\ref{fig:LQD_nubb}.

%
%
%
%

\subsection{``Generic'' LRPV\texorpdfstring{ ($\chi \to \ell q q, \nu q q$)}{}} \label{sec:Generic}

Finally we consider models in which the LSP decays 50\% of the time to a charged lepton and two light quarks and 50\% to a neutrino and two light quarks. Such branching ratios may be called ``generic'' for decays through $LQD$ operators due to $SU(2)$ symmetry, though as we have seen there are important exceptions e.g. in the case of higgsino LSPs or decays to the third generation.

\begin{figure}
		  \centering
        \subfigure[]{
                \includegraphics[width=3in]{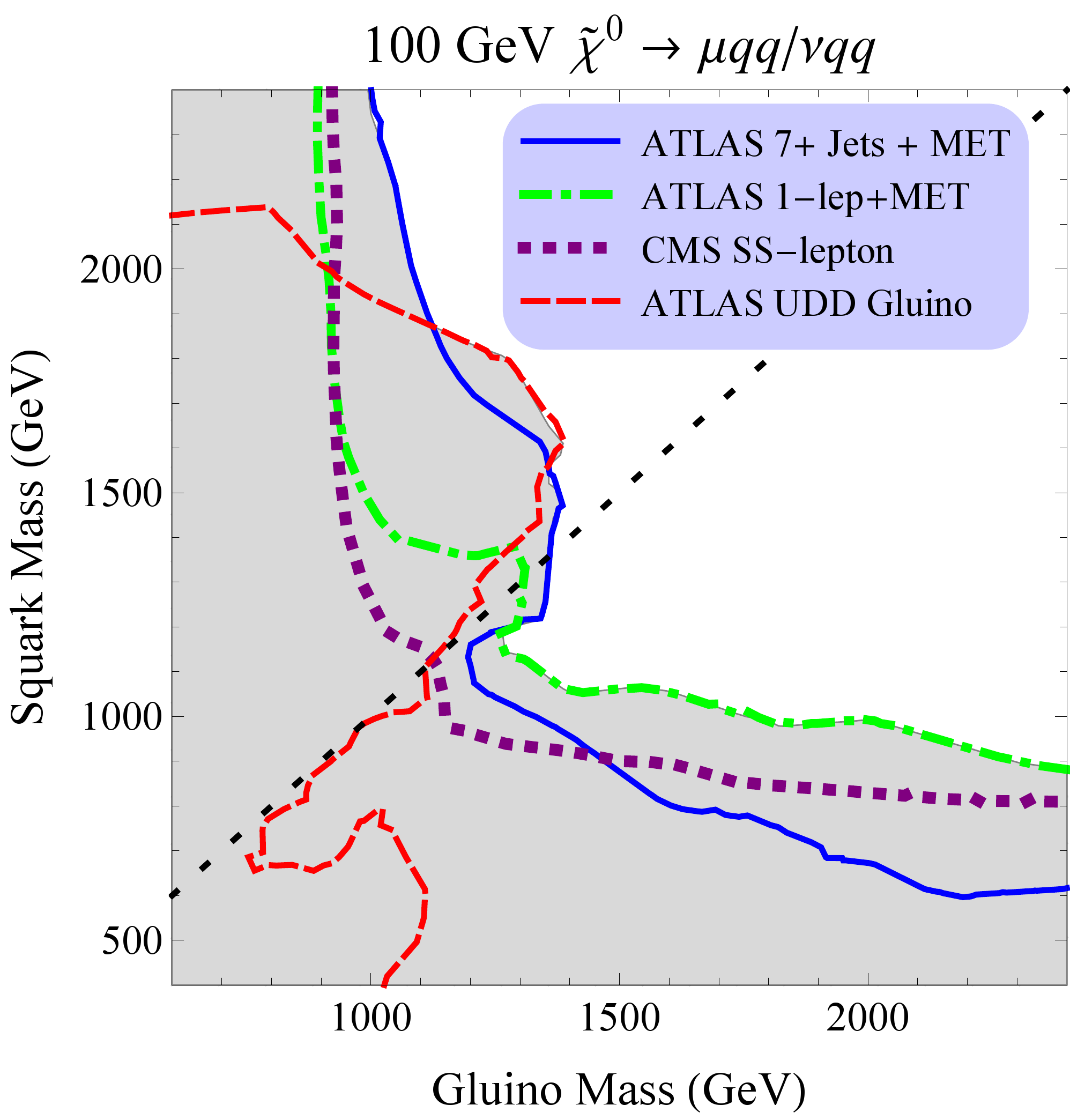}
                \label{fig:LQD_munu_100}
		  }
               \hspace{5mm} 
 	\subfigure[]{
                \includegraphics[width=3in]{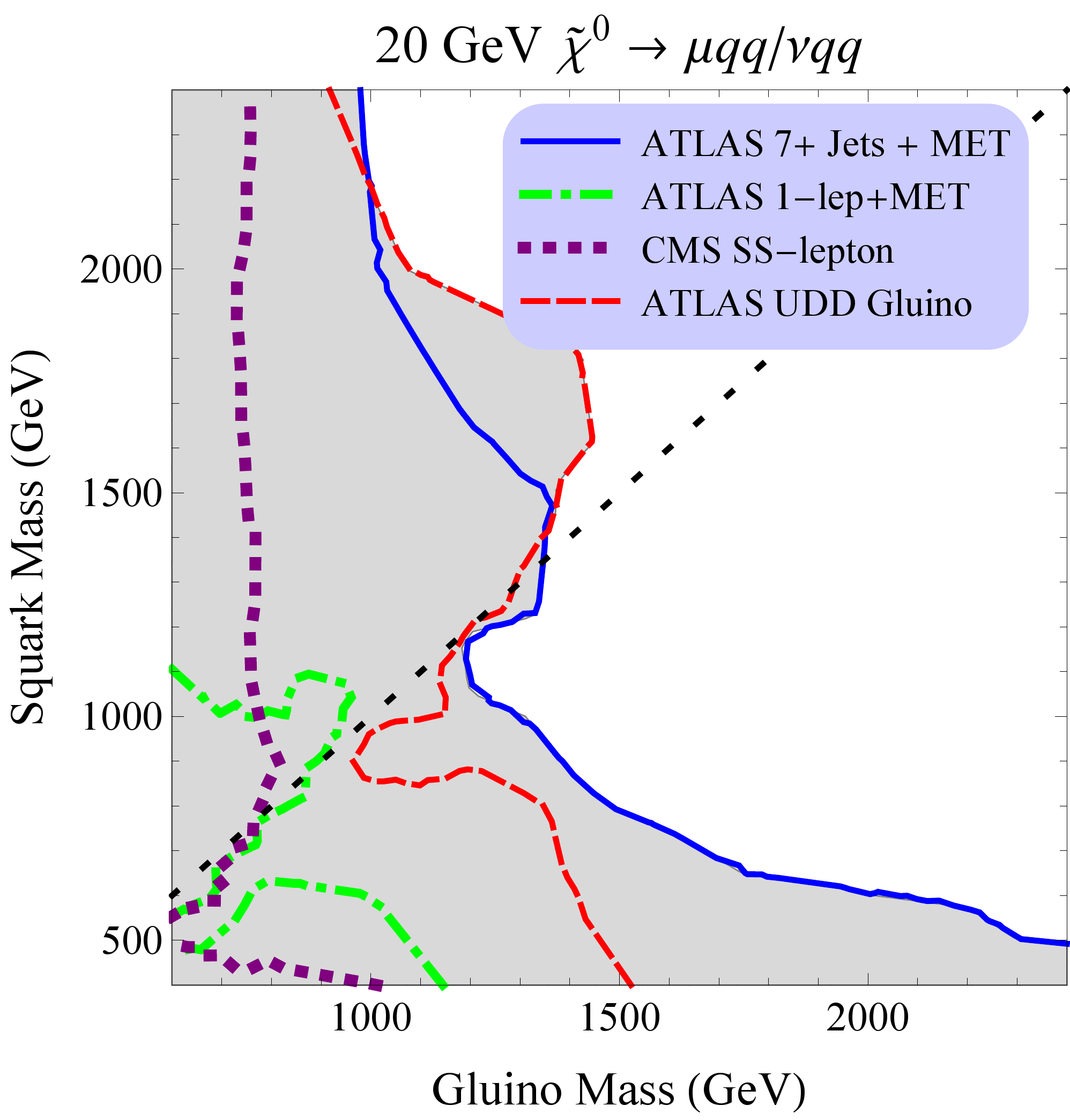}
                \label{fig:LQD_munu_20}
		  }
               \hspace{5mm} 
        \caption{Constraints on a model of the type depicted in figure~\ref{fig:LQD_lnu_spectrum}, with a gluino, degenerate squarks, and a neutralino decaying equally to $\mu q q$ and $\nu q q$. The relevant constraints in this case come from the ATLAS search for many jets and missing energy~\cite{ATLASMultijet20}, an ATLAS search for events with one lepton, jets and missing energy~\cite{ATLAS1lepMET20}, and the CMS search for same-sign leptons~\cite{CMSSSlep20}.}
        \label{fig:LQD_munu}
\end{figure}

\begin{figure}
		  \centering
        \subfigure[]{
                \includegraphics[width=3in]{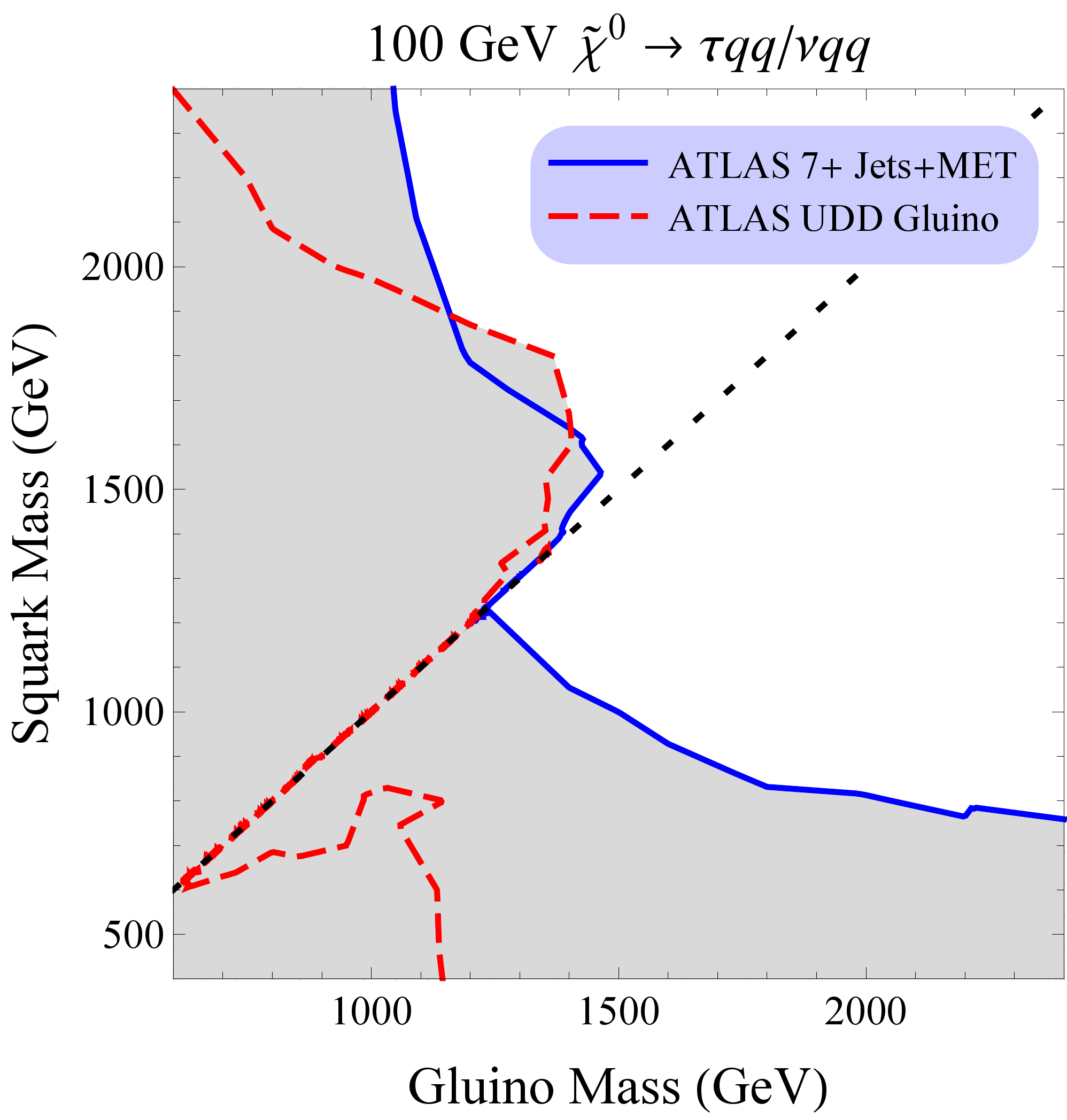}
                \label{fig:LQD_taunu_100}
		  }
               \hspace{5mm} 
 	\subfigure[]{
                \includegraphics[width=3in]{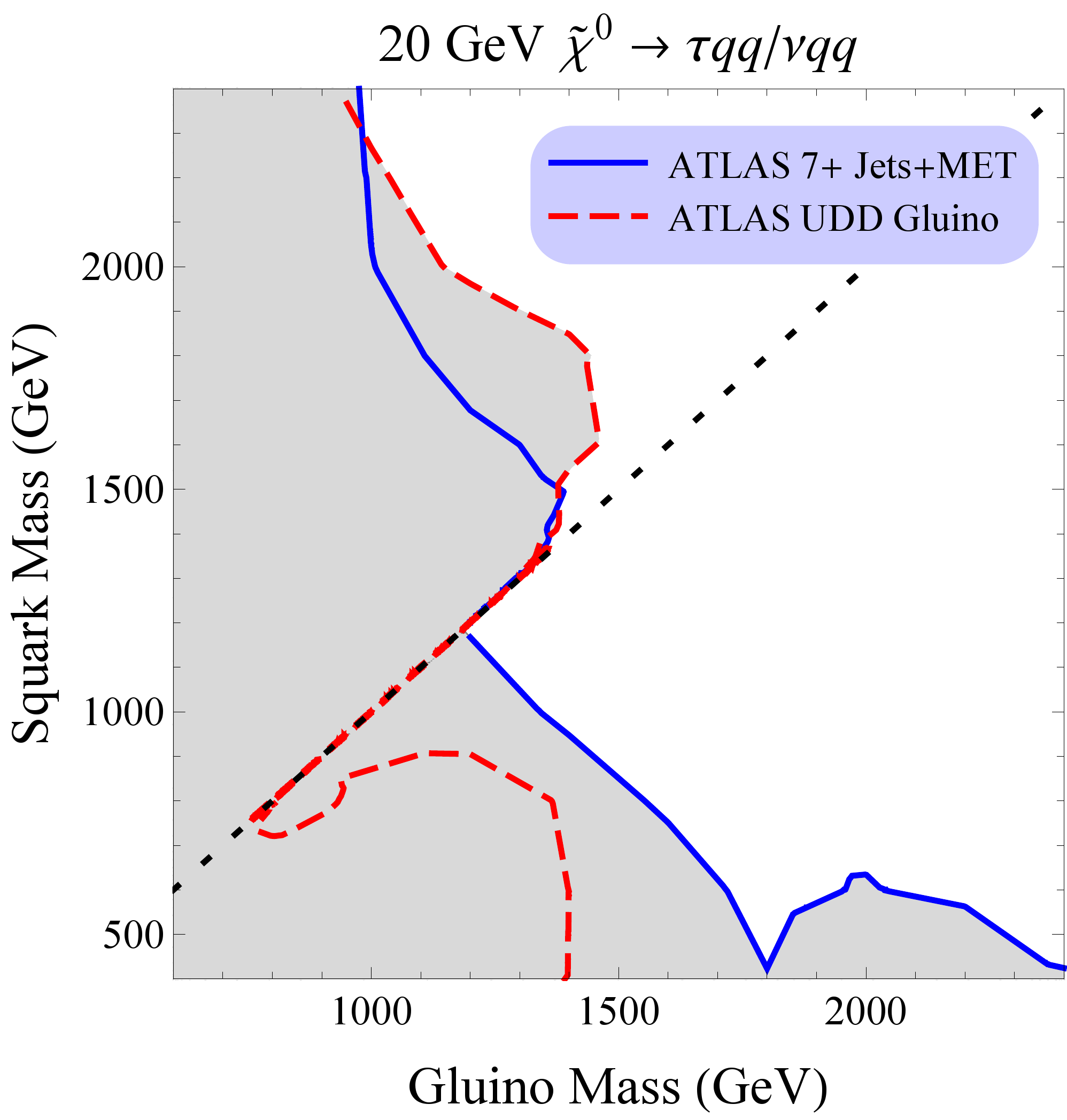}
                \label{fig:LQD_taunu_20}
		  }
               \hspace{5mm} 
        \caption{Constraints on a model as in figure~\ref{fig:LQD_munu}, but with the LSP decaying  equally to $\tau q q$ and $\nu q q$. The strongest bounds come from the ATLAS search in many jets and missing energy~\cite{ATLASMultijet20}.}
        \label{fig:LQD_taunu}
\end{figure}

In figure~\ref{fig:LQD_munu}, we show the constraints on the squark and gluino masses assuming the LSP decays to the final states $\mu q q$ and $\nu q q$ with equal branching ratios, for two different values of LSP mass. The strongest constraints arise from the ATLAS searches for many (7+) jets and missing energy~\cite{ATLASMultijet20} and for 1-lepton, jets and missing energy~\cite{ATLAS1lepMET20}. The constraints from the CMS search for same-sign leptons~\cite{CMSSSlep20} are generally weaker than the ATLAS 1-lepton + MET search. For a 100 GeV LSP the 1-lepton search places a better bound than the many-jet search on the squark mass, since squark production produces fewer jets than gluino production. As we saw previously, for a 20 GeV LSP the searches selecting leptons lose sensitivity as the leptons are no longer isolated.

Similarly, in figure~\ref{fig:LQD_taunu} we show the constraints for LSP decays to $\tau q q$ and $\nu q q$. Here the ATLAS many-jet + MET search places the strongest bounds throughout. Again, we do not claim any constraints from analyses which reconstruct hadronic taus.

In these models the squarks and gluinos are forced to be heavier than in the RPV scenarios without direct production of neutrinos. Once more, collimation of the LSP decay products helps alleviate the bounds.

\section{Conclusions} \label{sec:Conclusions}

\begin{figure}
\hspace*{20mm}
\includegraphics[width=6in]{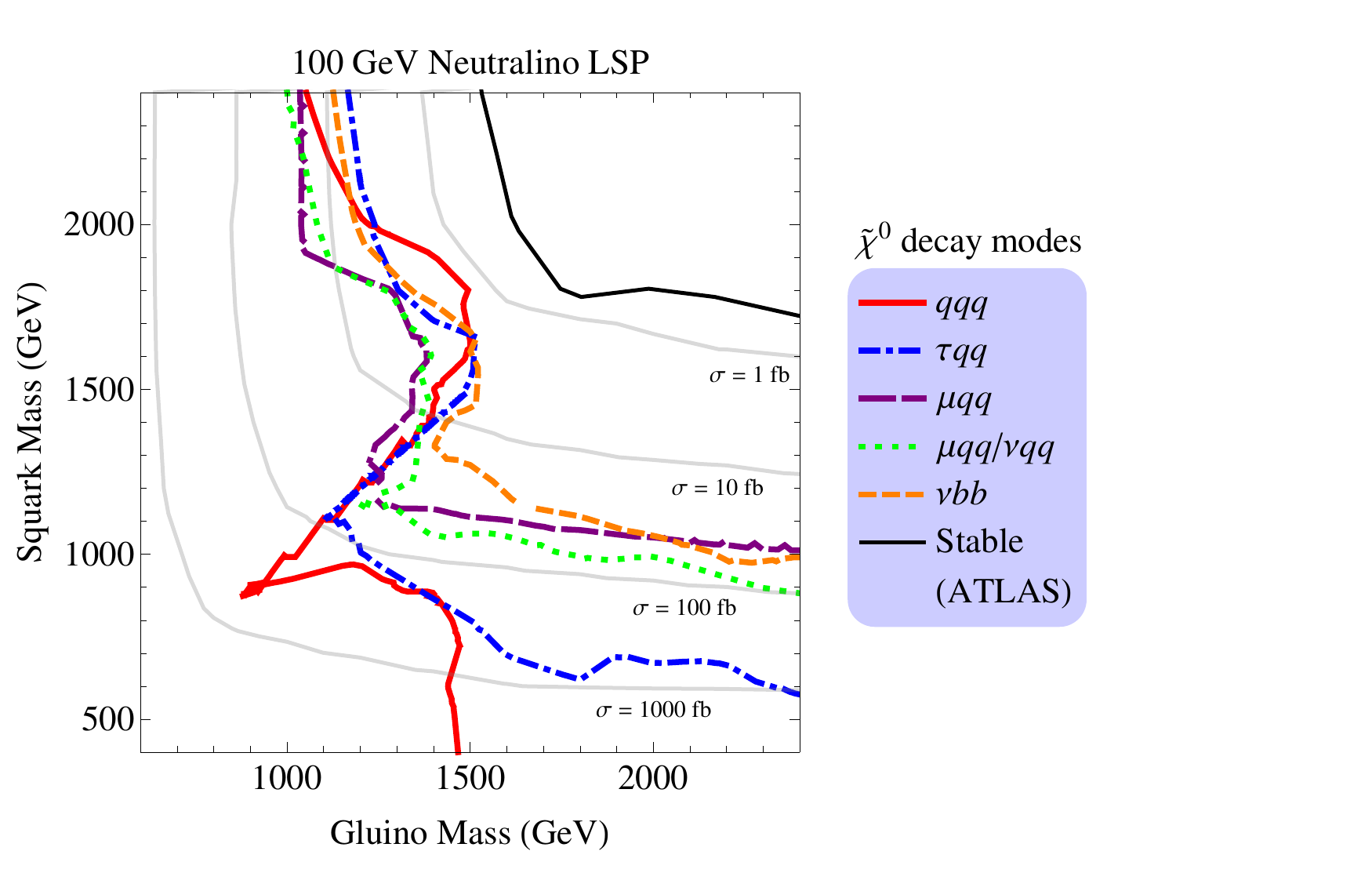}
\caption{Summary of constraints on models with squarks and gluinos decaying to a 100 GeV neutralino LSP (as in figure~\ref{fig:Spectra}). Each colored contour represents the combined constraint from all searches for a particular LSP decay mode. The solid black contour indicates the official ATLAS constraint on a model with gluinos, first and second generation squarks and a stable massless neutralino LSP, from a search in final states with 2-6 jets and missing energy~\cite{ATLASJetsMET20}. 
}         
\label{fig:Combined}
\end{figure}

\begin{figure}
\hspace*{20mm}
\includegraphics[width=6in]{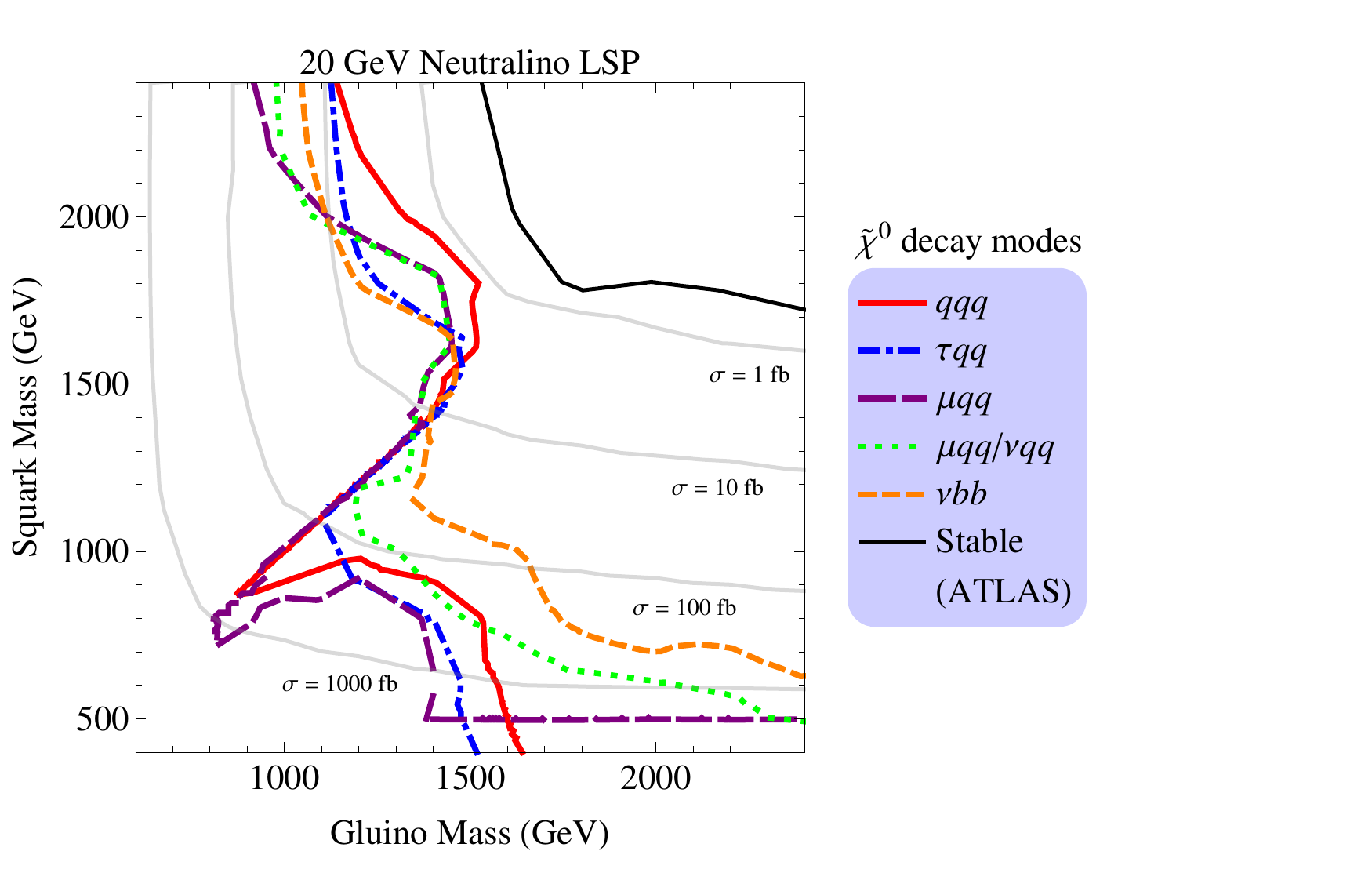}
\caption{As in figure~\ref{fig:Combined}, summary of constraints on models with squarks and gluinos decaying to a 20 GeV neutralino LSP. Each colored contour represents the combined constraint from all searches for a particular LSP decay mode, and the solid black contour indicates the constraint from  ATLAS~\cite{ATLASJetsMET20} for a stable massless LSP.
}         
\label{fig:Combined20}
\end{figure}

We have explored a variety of $R$-parity violating SUSY models with an eye towards hiding large SUSY cross-sections from current LHC searches. To this end we considered models of the form depicted in figure~\ref{fig:Spectra}, with a gluino, all three generations of squarks, and a single neutralino or nearly degenerate electroweakino multiplet. Such spectra were chosen to eliminate signatures such as gauge or Higgs bosons from cascade decays, leaving only hard jets from the gluino/squark decays plus the products of the RPV neutralino decay. Our results are summarized in figures~\ref{fig:Combined} and~\ref{fig:Combined20}, which show the constraints on the squark and gluino masses for several different LSP decay modes and two different LSP masses (20 GeV and 100 GeV), as well as for a stable massless LSP (from an ATLAS search in jets and MET~\cite{ATLASJetsMET20}). Superimposed on the same plane are contours of the total colored superpartner production cross-section; note that certain models allow for SUSY cross-sections as high as $\sim \pb$, implying several thousand new physics events in the existing 8 TeV LHC data. Our results challenge some of the conventional wisdom about RPV at colliders and suggest new collider searches that could have discovery potential even in the 8 TeV data.


The first scenario we considered, baryonic RPV (section~\ref{sec:Baryonic}), has long been thought to completely hide SUSY at colliders due to its completely hadronic signal. Recent searches, particularly~\cite{ATLASUDDGluino}, however have now constrained gluino production in this model, and we have shown (figure~\ref{fig:UDD}) that these bounds are particularly strong when the squarks are not decoupled, extending to $1.5 \TeV$ if the squarks are lighter than $\sim 1 \TeV$. The rapid transition to these bounds from no prior constraints at all was made possible by a simple new search strategy of selecting many (6+) jets with very high $p_T$. Since $R$-parity violation necessarily results in a high-multiplicity final state, selecting events with many jets appears to be an extremely powerful strategy, in some sense replacing the missing transverse momentum cuts used to search for $R$-parity conserving SUSY. Use of jet substructure techniques, as proposed in ~\cite{Cohen:2012yc, Hedri:2013pvl} and implemented to some extent in the ATLAS multijet+ MET search~\cite{ATLASMultijet20}, could help further distinguish SUSY signal events from QCD backgrounds.

We also considered a variety of leptonic RPV models. These are generally \emph{but not always} more strongly constrained thanks to the many existing searches for charged leptons and/or missing energy.  In particular, we identified several LRPV scenarios which allowed light colored superpartners, sometimes with even weaker constraints than baryonic RPV.  Decays of the LSP to a neutrino are strongly constrained by searches requiring high jet multiplicity and small but statistically significant missing energy (figure~\ref{fig:LQD_nubb}). Interestingly, decays of the LSP purely to a charged lepton and two quarks are less constrained by existing searches, despite the high proportion of events with a same-sign lepton final state. We identified a number of leptonic RPV scenarios in which the bounds on squarks and gluinos can be comparable to or even weaker than those for baryonic RPV models:

\begin{itemize}
  \item If the LSP is very light, such that it is produced with high boost in the decays of the squarks and/or gluino. The LSP decay products then tend to merge into a jet, causing the lepton to fail isolation criteria. Then despite every SUSY event containing two leptons, most searches selecting leptons will have very low efficiency and the constraints can be even weaker than those on baryonic RPV (figure~\ref{fig:LQD_emuqq_20}). 
  \item If the LSP has a \emph{displaced} decay to an electron and two quarks, then the bounds from typical searches in leptonic final states do not apply (top plots in figure~\ref{fig:LQD_emuqq}). The only possibly relevant displaced vertex search is the CMS displaced dijet search~\cite{CMSDisplacedDijet} which selects for long-lived particles with two-body decays but could have some nonzero efficiency for this final state.
  \item If the LSP decays 100\% to a tau and two jets, as is favored for a higgsino, then the bounds from many jet + missing energy searches are only slightly stronger than those on baryonic RPV (figure~\ref{fig:LQD_tauqq_nat}).
\end{itemize}       

These ``gaps'' in the coverage of leptonic RPV suggest new or modified searches. One general avenue for progress would be to search for events with many jets and one or more leptons. This capitalizes on both of the distinctive features of this type of RPV, much as the multijet + MET search is powerful for final states with neutrinos. Searches of this type have been proposed previously in the literature~\cite{LSST, CoverageNatural} in the context of models with high jet multiplicity and leptons from $W$'s, $Z$'s or tops; our models represent another qualitatively different scenario that could be probed by this type of search. The specific MC studies in these works consider selecting events with many jets, one lepton and either moderate MET or $b$-jets; to cover the models we discussed one could relax any MET or $b$-tag cuts but instead look for events with two (possibly same-sign) leptons. A search selecting taus (without requiring missing energy as existing analyses do), though more challenging, would target the highly motivated scenario of a higgsino LSP decaying through the $LQD$ operator.    

This strategy alone will not succeed in the case of highly boosted LSPs, where the efficiency for leptons to pass isolation requirements can approach zero. It would be interesting to explore the possibilities for searches which relax the standard lepton isolation criteria. This has already been done in searches for ``lepton jets", sprays of collimated leptons (with no overlapping hadronic particles)~\cite{ATLASLeptonJet,CMSLeptonJet}. Allowing leptons to overlap with hadronic energy introduces large backgrounds, including pure QCD scattering; however the multijet searches~\cite{ATLASUDDGluino} demonstrate that QCD backgrounds can be greatly attenuated by requiring many hard jets. It seems likely that searches selecting for non-isolated leptons plus many jets and/or high $H_T$ would have greater sensitivity to certain models than purely hadronic searches. Further assessing this claim requires careful study of the Standard Model backgrounds for non-isolated leptons. Note that many models motivating lepton jet searches~\cite{ArkaniHamed:2008qp, Baumgart:2009tn, Cheung:2009su} will also give rise to collimated showers including both hadrons and leptons in much of their parameter space, further motivating this type of search.    

We have seen that displaced decays can nullify most searches for new physics which rely on leptons or $b$-jets; however displaced vertices offer a very distinctive signature in their own right. Currently, there is a gap in the official coverage in the case of three-body displaced decays which do not produce muons (e.g., a decay to three quarks or an electron and two quarks). A CMS search for displaced dijets~\cite{CMSDisplacedDijet} could have sensitivity in this case depending on how restrictively they select for two-body decays. A similar analysis relaxing the assumption of a two-body decay could close all scenarios for displaced neutralino decays through RPV. (Decays through $LLE$ are constrained by a search for displaced dileptons~\cite{CMSDisplacedDilepton}).  

In all scenarios we considered the gluino is constrained to be heavier than $\sim900 \GeV$ (similar to the conclusions of~\cite{CoverageNatural}, although we considered different models). This lower bound is only achieved when the squarks are either very heavy or just lighter than the gluino; if the squarks are much lighter then the gluino bound can extend as high as $\sim1500 \GeV$. It remains possible, however, to have very light squarks in certain models, such as baryonic RPV (figure~\ref{fig:UDD}) or with leptonic RPV with boosted decays (figure~\ref{fig:LQD_emuqq_20}) These models survive since squark decay typically produces fewer jets than gluino decay, rendering multijet searches much less effective. Probing these squark-only models remains an open experimental challenge. In the case of boosted LSPs decaying to leptons, a search for non-isolated leptons as suggested above could have sensitivity. Scenarios with highly boosted LSPs also allow for squarks to be reconstructed as dijet resonances if no neutrinos are produced in the LSP decay. As we discussed, existing searches for paired dijet resonances~\cite{CMSDijetResonance} are only directly sensitive if the squarks are degenerate to within a few percent; however even if the squarks have significant mass splittings they could still give rise to distinct features in the paired dijet mass spectrum, such as broad bumps or multiple smaller peaks. It may be possible to constrain these more general features with new analyses, and/or to make use of paired dijet mass information in searches selecting for non-isolated leptons.

Our results have implications for questions of SUSY naturalness and model-building. We have shown that even in the most optimistic baryonic RPV scenarios, there are significant bounds in the squark-gluino mass plane which challenge naturalness, especially if all the squarks are approximately degenerate. Some leptonic RPV scenarios appear to rival baryonic RPV in terms of minimizing fine-tuning, in particular the scenario of a higgsino LSP decaying to a tau and two jets (section~\ref{sec:Tau-rich}). Although the collider phenomenology of this scenario resembles that of baryonic RPV, the implications for precision physics and cosmology are radically different. In particular, baryonic RPV models imply the washout of any cosmological baryon asymmetry down to very low temperatures if the RPV coupling is large enough to cause prompt decays at colliders~\cite{Baryonecrosis}, forcing baryogenesis to occur at very low temperatures. In contrast, leptonic RPV allows for baryogenesis any time during or after the electroweak phase transition, once sphalerons no longer equilibrate $B + L$ to zero.  

Throughout this work we have focused on simplified models, assuming minimal spectra and a few dominant couplings. This allowed us to more easily identify the most interesting possibilities for collider physics. It remains an open problem to construct more complete models that realize the phenomenology we have considered. Recently there has been considerable interest in building models of baryonic RPV, motivated by its assumed ability to hide low-energy SUSY; our results indicate that the study of leptonic RPV can be similarly motivated. More complete models could suggest correlated observables at colliders as well as in flavor and neutrino physics.

Finally we point out that, though all of the models we have examined have been couched in the language of supersymmetry, our observations on collider phenomenology apply generally. We have shown that there is room to improve the LHC constraints for models with many hard jets plus leptons, displaced vertices, boosted particles decaying to leptons, and/or tau-rich final states. We have argued that searching for these signatures would help close windows for light superpartners and deepen the crisis of electroweak naturalness. However, such signals are far from being unique to supersymmetry, and searching for them could also reveal unexpected new physics. The LHC has accumulated a treasure trove of data at the frontiers of physics; it behooves us to use it to its full potential for discovery.

\section*{Acknowledgments}
We acknowledge useful conversations with Tim Cohen, Yanou Cui, Savas Dimopoulos, Jared Evans, Roberto Franceschini, Beate Heinemann, Kiel Howe,  Lawrence Lee,  Emanuel Strauss,  Matthew Walker, Wells Wulsin, and Yue Zhao. PWG acknowledges the support of NSF grant PHY-1316706, the Hellman Faculty Scholars program, and the Terman Fellowship. SR was supported by ERC grant BSMOXFORD no. 228169. P.S. acknowledges the support of the Maryland Center for Fundamental Physics, and NSF grants PHY-1315155 and PHY-1214000.

\appendix

\section{Details of Monte Carlo simulation and validation} \label{sec:Appendix}

In this appendix we provide additional details on our procedure for Monte Carlo simulation of collider events and the reinterpretation of collider searches, as well as comparisons of our results using these methods to the official limits on specific models presented by the ATLAS and CMS collaborations.

The tools we use for simulation and analysis are {\tt MadGraph5}~\cite{MadGraph}, {\tt Pythia6.4}~\cite{PYTHIA}, a modified version of {\tt PGS4}~\cite{PGS}, and {\tt NLL-fast}~\cite{Beenakker:2011fu}. We use {\tt MadGraph} to generate squark/gluino pair-production only, and pass the events to {\tt Pythia} to decay all particles (including superpartners) and to perform showering, hadronization, radiation, etc. All superpartner branching ratios for our simplified models are supplied to {\tt Pythia} in a decay table. Because {\tt Pythia} uses a flat matrix element for three-body decays, when computing efficiencies for LHC searches we weight each event by the matrix elements for the particular neutralino decay kinematics in the event, assuming in all cases that the neutralino decays through an off-shell slepton (rather than a squark). The {\tt Pythia} output is passed to a version of the detector simulation software {\tt PGS} with the following modifications from the standard {\tt PGS4}:

\begin{itemize}

\item Instead of using the default $b$-tagging, we have {\tt PGS} identify ``truth" $b$-jets in its output, and then apply the $b$-tag efficiencies as a function of $p_T$ as relevant for individual searches~\cite{Kiel}.

\item Instead of the default lepton isolation cuts for electrons and muons we use the criteria specified in the actual leptonic searches, i.e~\cite{ATLAS1lepMET20} for ATLAS and~\cite{CMSSSlep20} for CMS. 

\item For ATLAS searches we do NOT merge non-isolated muons with jets, i.e. we assume that jets are constructed purely from calorimeter depositions.    

\item We use the default ATLAS and CMS detector parameter cards, except for modifying the jet clustering algorithm to reflect that actually used by the collaborations, namely anti-$k_T$ with clustering radius $R = .4$ (.5) for ATLAS (CMS). 

\end{itemize}  

To derive bounds on our models from LHC searches we apply the cuts used in these analyses to the {\tt PGS} output to compute signal efficiencies at each model point. We derive expected signal event yields by multiplying these efficiencies by the appropriate production cross-sections for squarks and gluinos, which we obtain at NLL order using the code {\tt NLL-fast}. These expected event yields are compared to the 95\% exclusion limits for each signal selection; a model is considered excluded if it exceeds the 95\% limit in any signal region. 

In the following sections we describe the validation of our simulation pipeline for certain searches by comparison with official ATLAS/CMS results.

\subsection{ATLAS Multijet + MET} \label{sec:ATLASMultijet20_Validate}

\begin{figure}
\includegraphics[width=3in]{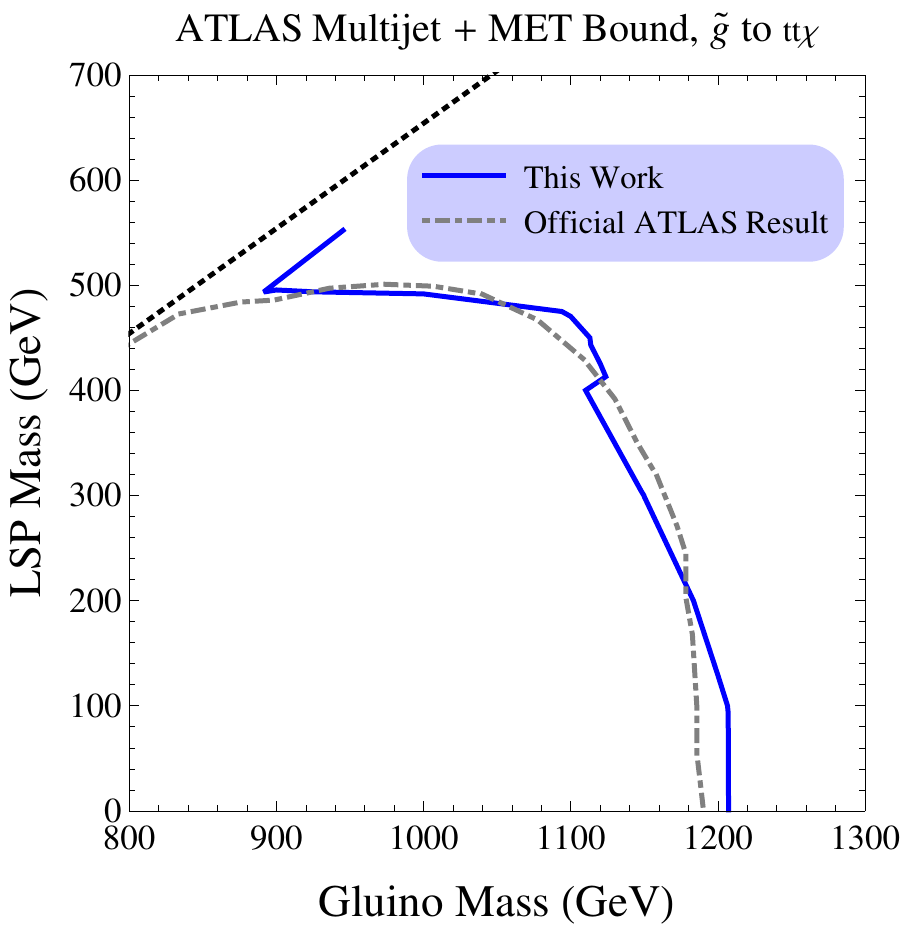}
\caption{Bounds from the ATLAS search in events with many (7+) jets and missing energy~\cite{ATLASMultijet20} on a model with gluinos decaying through heavy off-shell stops to top quarks and a stable neutralino LSP, as reported by the ATLAS collaboration (gray dash-dotted curve) and as determined using our simulation pipeline (solid blue curve). The black dotted line shows the kinematic boundary for the gluino decay, $m_{\tilde{g}} = 2 m_t + m_{\tilde{\chi}^0}$.
}         
\label{fig:ATLASMultijet20_Validation}
\end{figure}

To validate our methods for the ATLAS search in many jets and missing energy~\cite{ATLASMultijet20}, we simulate pair-production of gluinos decaying 100\% through off-shell stops to two tops and a stable neutralino LSP, and compare our bounds in the plane of  
gluino mass and LSP mass to the official ATLAS results for the same model. Both bounds are shown in figure~\ref{fig:ATLASMultijet20_Validation}. The exclusion curves agree very well up to the region near the kinematic boundary $m_{\tilde{g}} \sim 2 m_t + m_{\tilde{\chi}^0}$, which we did not sample finely. 

\subsection{ATLAS $UDD$ Gluino Search} \label{sec:ATLASUDDGluino_Validate}

\begin{table}[b]
\caption{Comparison of results from the official ATLAS MC and our pipeline for the expected signal counts in the ATLAS search for RPV gluinos. For each model ATLAS determines which signal selection places the optimal expected constraint; the signal counts shown here are for this optimal selection. We do not show models for which the optimal selection requires $b$-tagging, as we do not make use of these selections in placing bounds in this work.}
\begin{center}
\begin{tabular}{| l | c | c | c |}
\hline
\textbf{Signal Model} & \parbox[t]{3cm}{\textbf{Optimal \\ Selection}} & \parbox[t]{3cm}{\textbf{Expected Signal Counts \\ (ATLAS official)}}  & \parbox[t]{3cm}{\textbf{Expected Signal Counts \\ (Our MC)}}  \\ \hline
$\tilde{g}$ LSP, $m_{\tilde{g}} = 500 \GeV$ & 7j $p_T > 120 \GeV$& $600 \pm 230$ & 630 \\ \hline
$\tilde{g}$ LSP, $m_{\tilde{g}} = 600 \GeV$  & 7j $p_T > 120 \GeV$ & $410 \pm 100$ & 470 \\ \hline
$\tilde{g}$ LSP, $m_{\tilde{g}} = 800 \GeV$  & 7j $p_T > 180 \GeV$ & $13 \pm 4$ & 12 \\ \hline
$\tilde{g}$ LSP, $m_{\tilde{g}} = 1000 \GeV$ & 7j $p_T > 180 \GeV$ & $6.8 \pm 2.3$ & 7.6 \\ \hline
$\tilde{g}$ LSP, $m_{\tilde{g}} = 1200 \GeV$  & 7j $p_T > 180 \GeV$ & $2.7 \pm .5$ & 2.8 \\ \hline
 $\tilde{\chi}$ LSP, $m_{\tilde{g}} = 400 \GeV$, $m_{\tilde{\chi}} = 50 \GeV$ & 7j $p_T > 100 \GeV$ & $ 1400\pm800 $ & 1840 \\\hline
 $\tilde{\chi}$ LSP, $m_{\tilde{g}} = 400 \GeV$, $m_{\tilde{\chi}} = 300 \GeV$ & 7j $p_T > 80 \GeV$ & $ 9000\pm4000 $ & 13000 \\ \hline
 $\tilde{\chi}$ LSP, $m_{\tilde{g}} = 600 \GeV$, $m_{\tilde{\chi}} = 300 \GeV$ & 7j $p_T > 100 \GeV$ & $ 1700\pm900 $ & 2530 \\ \hline
 $\tilde{\chi}$ LSP, $m_{\tilde{g}} = 800 \GeV$, $m_{\tilde{\chi}} = 300 \GeV$ & 7j $p_T > 120 \GeV$ & $ 380\pm90 $ & 346 \\ \hline
 $\tilde{\chi}$ LSP, $m_{\tilde{g}} = 1000 \GeV$, $m_{\tilde{\chi}} = 50 \GeV$ & 6j $p_T > 180 \GeV$ & $ 40\pm6 $ & 39 \\ \hline
 $\tilde{\chi}$ LSP, $m_{\tilde{g}} = 1000 \GeV$, $m_{\tilde{\chi}} = 300 \GeV$ & 7j $p_T > 140 \GeV$ & $ 50\pm13 $ & 58 \\ \hline
 $\tilde{\chi}$ LSP, $m_{\tilde{g}} = 1000 \GeV$, $m_{\tilde{\chi}} = 600 \GeV$ & 7j $p_T > 180 \GeV$ & $ 10\pm5 $ & 14 \\ \hline
 $\tilde{\chi}$ LSP, $m_{\tilde{g}} = 1200 \GeV$, $m_{\tilde{\chi}} = 50 \GeV$ & 7j $p_T > 180 \GeV$ & $ 1.9\pm1.0 $ & 1.8 \\ \hline
 $\tilde{\chi}$ LSP, $m_{\tilde{g}} = 1200 \GeV$, $m_{\tilde{\chi}} = 300 \GeV$ & 7j $p_T > 180 \GeV$ & $ 3.2\pm1.4 $ & 4.7 \\ \hline
 $\tilde{\chi}$ LSP, $m_{\tilde{g}} = 1200 \GeV$, $m_{\tilde{\chi}} = 600 \GeV$ & 7j $p_T > 140 \GeV$ & $ 28\pm4 $ & 32 \\ \hline
\end{tabular}
\end{center}
\label{tab:ATLASRPVGluino20_Validation}
\end{table}

The ATLAS search for gluinos decaying through the $UDD$ operator~\cite{ATLASUDDGluino} has many non-exclusive signal regions selecting events with 6-7+ jets above various $p_T$ cuts, sometimes with $b$-tagging requirements. For every model point this analysis first determines with a blind analysis which signal region is \emph{expected} to give the strongest constraint, and then places a bound on the cross-section for that model point using the observed counts in that signal region alone. 



As discussed in section~\ref{sec:Baryonic}, we are only able to model decays through $UDD$ operators by introducing ``fake quarks", very light particles with boosted decays that look like jets at colliders. We justify this as an accurate representation of $UDD$ decays by reproducing the signal efficiencies reported in~\cite{ATLASUDDGluino} for various models and signal regions. For various model points, ATLAS reports which signal region they identified as giving the strongest expected constraint and gives the expected number of signal events in that signal region, with an error bar (second and third columns in table~\ref{tab:ATLASRPVGluino20_Validation}). We simulate each of these model points using our methods to obtain our estimate of the expected number of signal events (fourth column in table~\ref{tab:ATLASRPVGluino20_Validation}). Our results agree with those of ATLAS well within their quoted errors (which are dominated by limited Monte Carlo statistics~\cite{Larry}), suggesting that our modeling of $UDD$ decays captures the essential behavior.

The background predictions for the 7-jet signal regions are obtained from 5-jet control regions, which are assumed to have negligible signal contamination. We verified that for the allowed SUSY models we consider the signal contribution is $< 10\%$ of the total rate for the 5-jet control regions, comparable to the error in the MC predictions for the Standard Model background. Therefore these models are not constrained by the 5-jet data, nor do they significantly alter the background predictions from the 5-jet to 7-jet extrapolation.



\subsection{CMS Same-sign Lepton} \label{sec:CMSSSlep20_Validate}

\begin{figure}
\includegraphics[width=3in]{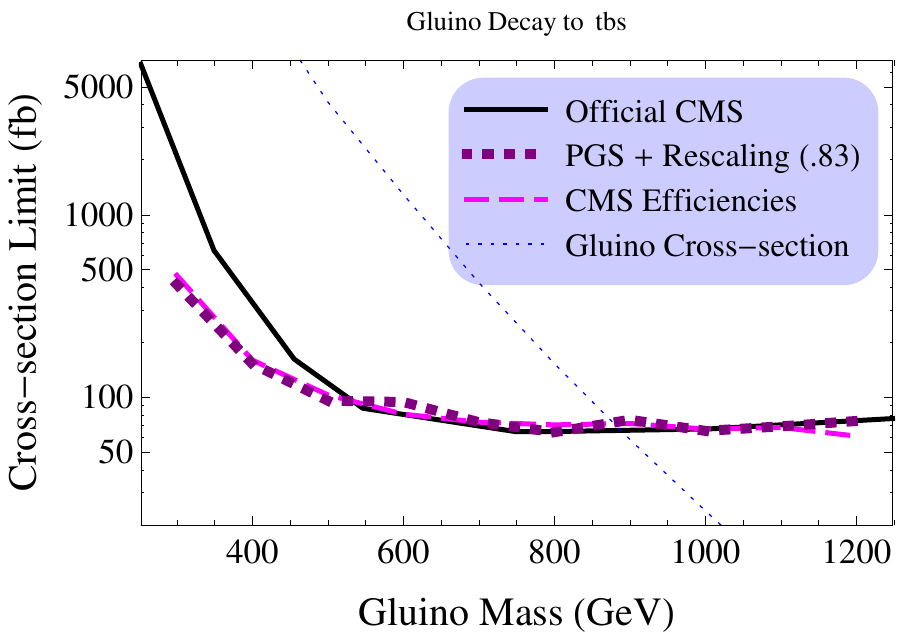}
\caption{Bounds from the CMS same-sign leptons search~\cite{CMSSSlep20} on a model with pair-produced gluinos each decaying to a top, bottom and strange quark, using the ``RPV2'' selection. The solid black curve is the official CMS result for the limit on the production cross-section as a function of gluino mass, while the dashed blue curve gives the actual gluino pair-production cross-section. The other two curves are bounds we derived by two methods described in the text: using {\tt PGS} to reconstruct all objects and then rescaling the efficiency by a constant factor of .83 (dotted purple), and applying the functions provided by CMS for reconstruction efficiency to the event-generator output (dashed magenta).   
}         
\label{fig:tbs}
\end{figure}

The CMS search in final states with same-sign leptons~\cite{CMSSSlep20} also defines many signal regions, most of which require missing energy and/or $b$-jets and are therefore not optimal for placing bounds on the RPV scenarios we consider. The most relevant signal region for our purposes is the ``RPV0" selection, requiring only two same-sign leptons and scalar sum of hadronic energy $H_T > 500 \GeV$. Unfortunately, we are unable to directly validate our results for the signal region, as CMS does not present any model interpretation making use of this region. We instead validate the similar selection ``RPV2", which requires two or more $b$-tagged jets in addition to the requirements or RPV0. This is used by CMS to place bounds on a model with pair-produced gluinos each decaying to a top, bottom and strange quark. The CMS constraints on the total gluino production cross-section in this model are shown in solid black in figure~\ref{fig:tbs},  and the gluino pair-production cross-section is in dashed blue.  

We attempted to reproduce the CMS bounds on this model using two methods. The first approach is to simulate the signal using the modified version of {\tt PGS} described above, and then allow for an overall constant rescaling of the signal efficiency. The rescaling is meant to account for the difference between the actual CMS lepton efficiencies and the {\tt PGS} efficiencies; a constant rescaling should suffice at high gluino mass since the lepton efficiency plateaus at high $p_T$. We find that rescaling the efficiency by a factor of .83 results in a bound that agrees with the CMS result for high gluino masses (dotted purple in figure~\ref{fig:tbs}). The second method is to determine the lepton efficiency and $H_T$ cut efficiency using the efficiency curves given in the appendix of~\cite{CMSSSlep20} applied to event-generator-level results (i.e., the {\tt PYTHIA} output). To determine the efficiency of the cut $H_T > 500 \GeV$ we use the efficiency function for $H_T > 400 \GeV$ with the midpoint $x_{1/2}$ and variance $\sigma^2$ both rescaled by $500/400 = 1.25$. This bound is shown in dashed magenta in figure~\ref{fig:tbs} (with no rescaling).

The bounds we derived using these methods agree well with the official CMS bounds for large gluino mass, although they diverge significantly for gluino masses below 500 GeV. The origin of this discrepancy (i.e., of the rapid rise in the cross-section limit reported by CMS as the gluino mass is lowered below 500 GeV) is unknown to us. However, we argue that for our purposes the agreement in the high gluino mass region is sufficient to validate our results, as the models we consider have squarks and gluinos $\gtrsim \TeV$, with the leptons carrying a similar fraction of the event energy. To derive the bounds presented in this work we use our {\tt PGS} simulation (so as to properly account for lepton isolation efficiencies) with the resulting signal efficiencies rescaled by a factor of .83 as was necessary here to match the CMS results. We have checked that this rescaling changes the bounds on squark/gluino masses by $\lesssim 50 \GeV$.


\subsection{ATLAS 1-lepton + MET} \label{sec:ATLAS1lepMET20_Validate}

The ATLAS lepton plus MET search in $20 \fb^{-1}$~\cite{ATLAS1lepMET20} defines a number of signal regions to target events with hard and soft leptons and varying amounts of hadronic activity and missing energy. For the models we consider the inclusive hard single-lepton signal regions are usually the most constraining. Model-independent bounds on the number of signal events are presented separately for the electron and muon channels. However, when interpreting their results in terms of specific models, the electron and muon channels are combined and information about the distributions of effective mass and MET in multiple bins is used rather than just counting events in the signal region; we do not attempt the latter due to lack of information on correlations between the expected background in different bins. 

\begin{figure}
\includegraphics[width=3in]{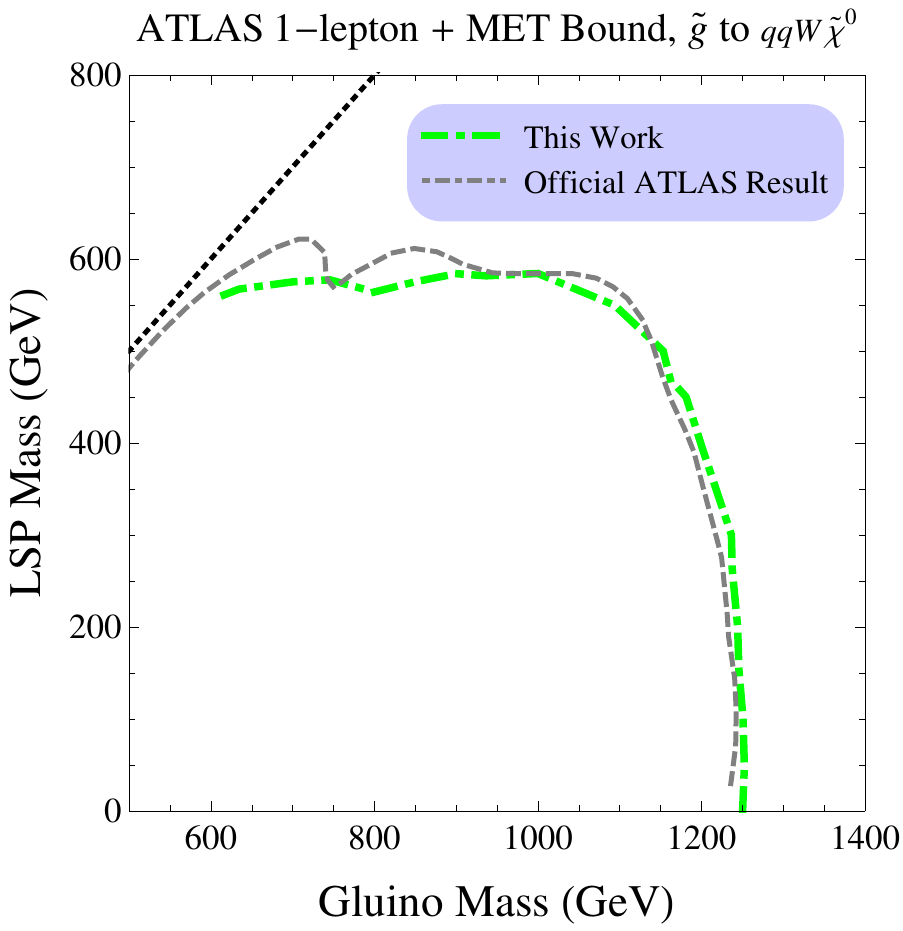}
\caption{Bounds from the ATLAS search in events with jets, leptons and missing energy~\cite{ATLAS1lepMET20} on a model with gluinos decaying to quarks and a chargino which then decays to a $W$ boson and stable neutralino LSP. The contours show the bounds as reported by the ATLAS collaboration (gray dashed curve) and as determined using our simulation pipeline using the model-independent signal regions (green dot-dashed curve). The black dotted line shows the kinematic boundary for the gluino decay, $m_{\tilde{g}} = m_{\tilde{\chi}^0}$.
}         
\label{fig:ATLAS1lepMET20_Validation}
\end{figure}

Although we are not able to exactly reproduce the ATLAS analysis, we nevertheless compare their bounds on a SUSY model against the results we obtain using the provided model-independent limits on the signal regions, the reasoning being that the high effective mass and high MET bins used to define these signal regions will be the most relevant for exclusion. We na\"{\i}vely combine the electron and muon channels by linearly adding the background uncertainties (i.e., assuming total correlation), and then compute bounds on the number of events in the combined channel by assuming a truncated Gaussian form for the expected background distribution. We compare our results for the case of what they call the ``one-step gluino'' model, in which the squarks are decoupled and the gluino decays to two light quarks and a chargino, with the chargino decaying to a $W$ boson and stable neutralino. The chargino mass is taken to be halfway between the gluino and LSP masses. Our results and the official ATLAS bounds are shown in figure~\ref{fig:ATLAS1lepMET20_Validation}, displaying good agreement.

\end{document}